\documentclass[12pt]{article}
\usepackage{amssymb}
\usepackage{amsmath}

\allowdisplaybreaks
\newtheorem{theorem}{Theorem}

\begin{document}

\title{Reducible second-class constraints of order $L$: an irreducible
approach}
\date{}
\author{C. Bizdadea\thanks{%
E-mail address: bizdadea@central.ucv.ro}, E. M. Cioroianu\thanks{%
E-mail address: manache@central.ucv.ro}, I. Negru\thanks{%
E-mail address: inegru@central.ucv.ro}, \and S. O. Saliu\thanks{%
E-mail address: osaliu@central.ucv.ro}, S. C. S\u{a}raru\thanks{%
E-mail address: scsararu@central.ucv.ro}, O. B\u{a}lu\c{s} \\
Faculty of Physics, University of Craiova\\
13 Al. I. Cuza Str., Craiova 200585, Romania}
\maketitle

\begin{abstract}
An irreducible canonical approach to second-class constraints reducible of
an arbitrary order is given. This method generalizes our previous results
from \cite{EPL,JPA} for first- and respectively second-order reducible
second-class constraints. The general procedure is illustrated on Abelian
gauge-fixed $p$-forms.
\end{abstract}

\section{Introduction}

The canonical approach to systems with reducible second-class constraints is
quite intricate, demanding a modification of the usual rules as the matrix
of the Poisson brackets among the constraint functions is not invertible.
Thus, it is necessary to isolate a set of independent constraints and then
construct the Dirac bracket \cite{1,2} with respect to this set. The split
of constraints may however lead to the loss of important symmetries, so it
should be avoided. As shown in \cite{3,4,5,6,7,9}, it is however possible to
construct the Dirac bracket in terms of a noninvertible matrix without
separating the independent constraint functions. A third possibility is to
substitute (by an appropriate extension of the phase-space) the reducible
second-class constraints with some irreducible ones, defined in the extended
phase-space, and further work with the Dirac bracket based on the
irreducible constraints. This idea, suggested in \cite{8} mainly in the
context of 2- and 3-form gauge fields, has been developed in a general
manner only for first- and respectively second-order reducible second-class
constraints \cite{EPL,JPA}. Other interesting contributions to reducible
second-class constrained systems (including the split involution formalism)
can be found in \cite%
{simon92MPLA,simon00IJMPA,simon96NPB,simon96PLB,simon99NPB}. The idea of
extending the phase-space is not new. It has been used previously, for
instance in the context of the conversion approach exposed in \cite{batalin}%
, where some supplementary variables are added in order to convert a set of
irreducible second-class constraints into a first-class one.

In this paper we give an irreducible approach to third-order reducible
second-class constraints and then generalize the results to an arbitrary
order of reducibility, $L$. Our strategy includes three main steps. First,
we express the Dirac bracket for the reducible system in terms of an
invertible matrix. Second, we construct an intermediate reducible
second-class system (of the same reducibility order like the original one)
on a larger phase-space and establish the (weak) equality between the
original Dirac bracket and that corresponding to the intermediate theory.
Third, we prove that there exists an irreducible second-class constraint set
equivalent to the intermediate one, such that the corresponding Dirac
brackets coincide (weakly). These three steps enforce the fact that the
fundamental Dirac brackets derived within the irreducible and original
reducible settings coincide (weakly). The equality between the fundamental
Dirac brackets associated with the original phase-space variables in the
reducible and respectively irreducible formulations has major implications
on the relationship between the reducible and irreducible systems: i) the
two systems exhibit the same number of physical degrees of freedom, which is
precisely the rank of the induced symplectic form (since the Dirac bracket
restricted to the constraint surface is determined by the inverse of the
induced symplectic form, see Theorem 2.5 from \cite{7}); ii) the physical
content of the two theories is the same from the perspective of quantization
as they display the same fundamental observables; iii) the original,
reducible system can be equivalently replaced with the irreducible one. It
is important to remark that the irreducible approach is useful mainly in
field theory because it does not spoil the important symmetries of the
original system, such as the spacetime locality of second-class field
theories.

The present paper is organized into six sections. In Section \ref{rev12} we
briefly review the procedure for second-class constraints that are reducible
of order one and respectively two. Sections \ref{third} and \ref{Lth} define
the `hard core' of the paper. We initially approach second-class constraints
reducible of order three in Section \ref{third} by implementing the three
main steps mentioned above, and then generalize these results to an
arbitrary order of reducibility in Section \ref{Lth}. In Section \ref{pforms}
we exemplify in detail the general procedure from Section \ref{Lth} on
gauge-fixed Abelian $p$-form gauge fields. Section \ref{conc} ends the paper
with the main conclusions.

\section{First- and second-order reducible second-class constraints: a brief
review\label{rev12}}

\subsection{Dirac bracket for first- and second-order reducible second-class
constraints}

We start with a system locally described by $N$ canonical pairs $%
z^{a}=\left( q^{i},p_{i}\right) $ and subject to the constraints
\begin{equation}
\chi _{\alpha _{0}}\left( z^{a}\right) \approx 0,\qquad \alpha _{0}=%
\overline{1,M_{0}}.  \label{1}
\end{equation}%
For simplicity, we take all the phase-space variables to be bosonic.
However, our analysis can be extended to fermionic degrees of freedom modulo
including some appropriate phase factors. We choose the scenario of systems
with a finite number of degrees of freedom only for notational simplicity,
but our approach is equally valid for field theories. In addition, we
presume that the functions $\chi _{\alpha _{0}}$ are not all independent,
but there exist some nonvanishing functions $Z_{\alpha _{1}}^{\alpha _{0}}$
such that
\begin{equation}
Z_{\alpha _{1}}^{\alpha _{0}}\chi _{\alpha _{0}}=0,\qquad \alpha _{1}=%
\overline{1,M_{1}}.  \label{2}
\end{equation}%
Moreover, we assume that the functions $Z_{\alpha _{1}}^{\alpha _{0}}$ are
all independent and (\ref{2}) are the only reducibility relations with
respect to the constraints (\ref{1}). These constraints are purely second
class if any maximal, independent set of $M_{0}-M_{1}$ constraint functions $%
\chi _{A}$ ($A=\overline{1,M_{0}-M_{1}}$) among the $\chi _{\alpha _{0}}$ is
such that the matrix
\begin{equation}
C_{AB}^{\left( 1\right) }=\left[ \chi _{A},\chi _{B}\right]  \label{3}
\end{equation}%
is invertible. Here and in the following the symbol $\left[ ,\right] $
denotes the Poisson bracket. In terms of independent constraints, the Dirac
bracket takes the form
\begin{equation}
\left[ F,G\right] ^{\left( 1\right) \ast }=\left[ F,G\right] -\left[ F,\chi
_{A}\right] M^{\left( 1\right) AB}\left[ \chi _{B},G\right] ,  \label{4}
\end{equation}%
where $M^{\left( 1\right) AB}C_{BC}^{\left( 1\right) }\approx \delta
_{C}^{A} $. In the previous relations we introduced an extra index, $\left(
1\right) $, having the role to emphasize that the Dirac bracket given in (%
\ref{4}) is based on a first-order reducible second-class constraint set. We
can rewrite the Dirac bracket expressed by (\ref{4}) without finding a
definite subset of independent second-class constraints as follows. We start
with the matrix
\begin{equation}
C_{\alpha _{0}\beta _{0}}^{\left( 1\right) }=\left[ \chi _{\alpha _{0}},\chi
_{\beta _{0}}\right] ,  \label{5}
\end{equation}%
which clearly is not invertible because
\begin{equation}
Z_{\alpha _{1}}^{\alpha _{0}}C_{\alpha _{0}\beta _{0}}^{\left( 1\right)
}\approx 0.  \label{6}
\end{equation}%
If $\bar{a}_{\alpha _{0}}^{\alpha _{1}}$ is a solution to the equation
\begin{equation}
\bar{a}_{\alpha _{0}}^{\alpha _{1}}Z_{\beta _{1}}^{\alpha _{0}}\approx
\delta _{\beta _{1}}^{\alpha _{1}},  \label{7}
\end{equation}%
then we can introduce a matrix \cite{6} of elements $M^{\left( 1\right)
\alpha _{0}\beta _{0}}$ through the relation
\begin{equation}
M^{\left( 1\right) \alpha _{0}\beta _{0}}C_{\beta _{0}\gamma _{0}}^{\left(
1\right) }\approx \delta _{\gamma _{0}}^{\alpha _{0}}-Z_{\alpha
_{1}}^{\alpha _{0}}\bar{a}_{\gamma _{0}}^{\alpha _{1}}\equiv d_{\gamma
_{0}}^{\alpha _{0}},  \label{8}
\end{equation}%
with $M^{\left( 1\right) \alpha _{0}\beta _{0}}=-M^{\left( 1\right) \beta
_{0}\alpha _{0}}$. Then, formula \cite{6}
\begin{equation}
\left[ F,G\right] ^{\left( 1\right) \ast }=\left[ F,G\right] -\left[ F,\chi
_{\alpha _{0}}\right] M^{\left( 1\right) \alpha _{0}\beta _{0}}\left[ \chi
_{\beta _{0}},G\right] ,  \label{10}
\end{equation}%
defines the same Dirac bracket like (\ref{4}) on the surface (\ref{1}). We
remark that there exist some ambiguities in defining the matrix of elements $%
M^{\left( 1\right) \alpha _{0}\beta _{0}}$ since if we make the
transformation
\begin{equation}
M^{\left( 1\right) \alpha _{0}\beta _{0}}\rightarrow M^{\left( 1\right)
\alpha _{0}\beta _{0}}+Z_{\alpha _{1}}^{\alpha _{0}}q^{\alpha _{1}\beta
_{1}}Z_{\beta _{1}}^{\beta _{0}},  \label{10wq}
\end{equation}%
with $q^{\alpha _{1}\beta _{1}}$ some completely antisymmetric functions,
then equation (\ref{8}) is still satisfied. Relations (\ref{6}) and (\ref{8}%
) show that%
\begin{equation}
\mathrm{rank}\left( d_{\gamma _{0}}^{\alpha _{0}}\right) \approx M_{0}-M_{1},
\label{10a}
\end{equation}%
which ensures the fact that the rank of the matrix of elements $M^{\left(
1\right) \alpha _{0}\beta _{0}}C_{\beta _{0}\gamma _{0}}^{\left( 1\right) }$
is equal to the number of independent second-class constraints in the
presence of the first-order reducibility.

Let us extend the previous construction to the case of second-order
reducible second-class constraints. This means that not all of the
first-order reducibility functions $Z_{\alpha _{1}}^{\alpha _{0}}$ are
independent. Beside the first-order reducibility relations (\ref{2}), there
appear also the second-order reducibility relations
\begin{equation}
Z_{\alpha _{2}}^{\alpha _{1}}Z_{\alpha _{1}}^{\alpha _{0}}\approx 0,\qquad
\alpha _{2}=\overline{1,M_{2}}.  \label{11x}
\end{equation}%
We will assume that the reducibility stops at order two, so all the
functions $Z_{\alpha _{2}}^{\alpha _{1}}$ are by hypothesis taken to be
independent. It is understood that the functions $Z_{\alpha _{2}}^{\alpha
_{1}}$ define a complete set of reducibility functions for $Z_{\alpha
_{1}}^{\alpha _{0}}$. In this situation, the number of independent
second-class constraints is equal to $M_{0}-M_{1}+M_{2}$. As a consequence,
we can work with a Dirac bracket of the type (\ref{4}), but in terms of $%
M_{0}-M_{1}+M_{2}$ independent functions $\chi _{A}$
\begin{equation}
\left[ F,G\right] ^{\left( 2\right) \ast }=\left[ F,G\right] -\left[ F,\chi
_{A}\right] M^{\left( 2\right) AB}\left[ \chi _{B},G\right] ,\qquad A=%
\overline{1,M_{0}-M_{1}+M_{2}},  \label{11b}
\end{equation}%
where $M^{\left( 2\right) AB}C_{BC}^{\left( 2\right) }\approx \delta
_{C}^{A} $, with $C_{AB}^{\left( 2\right) }=\left[ \chi _{A},\chi _{B}\right]
$. It is obvious that the matrix of elements
\begin{equation}
C_{\alpha _{0}\beta _{0}}^{\left( 2\right) }=\left[ \chi _{\alpha _{0}},\chi
_{\beta _{0}}\right]  \label{11d}
\end{equation}%
satisfies the relations
\begin{equation}
Z_{\alpha _{1}}^{\alpha _{0}}C_{\alpha _{0}\beta _{0}}^{\left( 2\right)
}\approx 0,  \label{11e}
\end{equation}%
so its rank is equal to $M_{0}-M_{1}+M_{2}$.

Let $\bar{A}_{\alpha _{1}}^{\alpha _{2}}$ be a solution of the equation
\begin{equation}
Z_{\beta _{2}}^{\alpha _{1}}\bar{A}_{\alpha _{1}}^{\alpha _{2}}\approx
\delta _{\beta _{2}}^{\alpha _{2}}  \label{a2}
\end{equation}%
and $\bar{\omega}_{\beta _{1}\gamma _{1}}=-\bar{\omega}_{\gamma _{1}\beta
_{1}}$ a solution to
\begin{equation}
Z_{\beta _{2}}^{\beta _{1}}\bar{\omega}_{\beta _{1}\gamma _{1}}\approx 0.
\label{a1}
\end{equation}%
We define an antisymmetric matrix, of elements $\hat{\omega}^{\alpha
_{1}\beta _{1}}$, through the relation
\begin{equation}
\hat{\omega}^{\alpha _{1}\beta _{1}}\bar{\omega}_{\beta _{1}\gamma
_{1}}\approx \delta _{\gamma _{1}}^{\alpha _{1}}-Z_{\alpha _{2}}^{\alpha
_{1}}\bar{A}_{\gamma _{1}}^{\alpha _{2}}\equiv D_{\gamma _{1}}^{\alpha _{1}}.
\label{a3}
\end{equation}%
Taking (\ref{a1}) into account, it results that $\hat{\omega}^{\alpha
_{1}\beta _{1}}$ contains some ambiguities, namely it is defined up to the
transformation
\begin{equation}
\hat{\omega}^{\alpha _{1}\beta _{1}}\rightarrow \hat{\omega}^{\alpha
_{1}\beta _{1}}+Z_{\alpha _{2}}^{\alpha _{1}}q^{\alpha _{2}\beta
_{2}}Z_{\beta _{2}}^{\beta _{1}},  \label{az}
\end{equation}%
with $q^{\alpha _{2}\beta _{2}}$ some arbitrary, antisymmetric functions. On
the other hand, simple computation shows that the matrix of elements $%
D_{\gamma _{1}}^{\alpha _{1}}$ satisfies the properties
\begin{eqnarray}
\bar{A}_{\alpha _{1}}^{\alpha _{2}}D_{\gamma _{1}}^{\alpha _{1}} &\approx
&0,\qquad Z_{\gamma _{2}}^{\gamma _{1}}D_{\gamma _{1}}^{\alpha _{1}}\approx
0,  \label{ax} \\
Z_{\alpha _{1}}^{\alpha _{0}}D_{\gamma _{1}}^{\alpha _{1}} &\approx
&Z_{\gamma _{1}}^{\alpha _{0}},\qquad D_{\gamma _{1}}^{\alpha
_{1}}D_{\lambda _{1}}^{\gamma _{1}}\approx D_{\lambda _{1}}^{\alpha _{1}}.
\label{ay}
\end{eqnarray}%
Based on the latter formula from (\ref{ax}), we infer an alternative
expression for $D_{\gamma _{1}}^{\alpha _{1}}$, namely
\begin{equation}
D_{\gamma _{1}}^{\alpha _{1}}\approx \bar{A}_{\alpha _{0}}^{\alpha
_{1}}Z_{\gamma _{1}}^{\alpha _{0}},  \label{1qa}
\end{equation}%
for some functions $\bar{A}_{\alpha _{0}}^{\alpha _{1}}$. From the former
relation in (\ref{ay}) and (\ref{1qa}) we deduce that
\begin{equation}
Z_{\gamma _{1}}^{\gamma _{0}}D_{\gamma _{0}}^{\alpha _{0}}\approx 0,
\label{12b}
\end{equation}%
where
\begin{equation}
D_{\gamma _{0}}^{\alpha _{0}}\approx \delta _{\gamma _{0}}^{\alpha
_{0}}-Z_{\alpha _{1}}^{\alpha _{0}}\bar{A}_{\gamma _{0}}^{\alpha _{1}}.
\label{11f}
\end{equation}%
At this stage, we can rewrite the Dirac bracket given in (\ref{11b}) without
separating a specific subset of independent constraints. In view of this, we
introduce an antisymmetric matrix, of elements $M^{\left( 2\right) \alpha
_{0}\beta _{0}}$, through the relation
\begin{equation}
M^{\left( 2\right) \alpha _{0}\beta _{0}}C_{\beta _{0}\gamma _{0}}^{\left(
2\right) }\approx D_{\gamma _{0}}^{\alpha _{0}},  \label{11c}
\end{equation}%
such that formula
\begin{equation}
\left[ F,G\right] ^{\left( 2\right) \ast }=\left[ F,G\right] -\left[ F,\chi
_{\alpha _{0}}\right] M^{\left( 2\right) \alpha _{0}\beta _{0}}\left[ \chi
_{\beta _{0}},G\right]  \label{14q}
\end{equation}%
defines the same Dirac bracket like (\ref{11b}) on the surface (\ref{1}). It
is simple to see that $M^{\left( 2\right) \alpha _{0}\beta _{0}}$ also
contains some ambiguities, being defined up to the transformation
\begin{equation}
M^{\left( 2\right) \alpha _{0}\beta _{0}}\rightarrow M^{\left( 2\right)
\alpha _{0}\beta _{0}}+Z_{\alpha _{1}}^{\alpha _{0}}\hat{q}^{\alpha
_{1}\beta _{1}}Z_{\beta _{1}}^{\beta _{0}},  \label{14r}
\end{equation}%
with $\hat{q}^{\alpha _{1}\beta _{1}}$ some antisymmetric, but otherwise
arbitrary functions. Relations (\ref{11x}) and (\ref{12b}) ensure that
\begin{equation}
\mathrm{rank}\left( D_{\gamma _{0}}^{\alpha _{0}}\right) \approx
M_{0}-M_{1}+M_{2},  \label{12a}
\end{equation}%
so the rank of the matrix of elements $M^{\left( 2\right) \alpha _{0}\beta
_{0}}C_{\beta _{0}\gamma _{0}}^{\left( 2\right) }$ is equal to the number of
independent second-class constraints also in the presence of the
second-order reducibility.

Direct manipulations emphasize that the Dirac bracket in each case, (\ref{10}%
) and (\ref{14q}) respectively, satisfies the relations%
\begin{equation}
\left[ \chi _{\alpha _{0}},G\right] ^{\left( 1,2\right) \ast }\approx 0,
\label{q12}
\end{equation}%
(where the index $(1)$ corresponds to (\ref{10}) and the index $(2)$ to (\ref%
{14q}) respectively), so the property $\left[ \chi _{\alpha _{0}},G\right]
^{\left( 1,2\right) \ast }=0$ (for any $G$) indeed holds on the surface of
first- or second-order reducible second-class constraints respectively. In
the meanwhile, each of the Dirac brackets (\ref{10}) or (\ref{14q})
satisfies the Jacobi identity, but only in the weak sense.

\subsection{Irreducible analysis of first- and second-order reducible
second-class constraints}

As it has been shown in \cite{EPL}, first-order reducible second-class
constraints can be approached in an irreducible manner. To this end, one
starts from the solution to equation (\ref{7})
\begin{equation}
\bar{a}_{\alpha _{0}}^{\alpha _{1}}=\bar{D}_{\gamma _{1}}^{\alpha
_{1}}a_{\alpha _{0}}^{\gamma _{1}},  \label{11m}
\end{equation}%
where $a_{\alpha _{0}}^{\gamma _{1}}$ are some functions chosen such that
\begin{equation}
\mathrm{rank}\left( Z_{\alpha _{1}}^{\alpha _{0}}a_{\alpha _{0}}^{\gamma
_{1}}\right) =M_{1}  \label{12}
\end{equation}%
and $\bar{D}_{\gamma _{1}}^{\beta _{1}}$ stands for the inverse of $%
Z_{\alpha _{1}}^{\alpha _{0}}a_{\alpha _{0}}^{\gamma _{1}}$. In order to
develop an irreducible approach, it is necessary to enlarge the original
phase-space with some new variables, $\left( Y_{\alpha _{1}}\right) _{\alpha
_{1}=\overline{1,M_{1}}}$, endowed with the Poisson brackets
\begin{equation}
\left[ Y_{\alpha _{1}},Y_{\beta _{1}}\right] =\Gamma _{\alpha _{1}\beta
_{1}},  \label{12xx}
\end{equation}%
where $\Gamma _{\alpha _{1}\beta _{1}}$ are the elements of an invertible,
antisymmetric matrix that may depend on the newly added variables.
Consequently, one constructs the constraints
\begin{equation}
\bar{\chi}_{\alpha _{0}}=\chi _{\alpha _{0}}+a_{\alpha _{0}}^{\alpha
_{1}}Y_{\alpha _{1}}\approx 0,  \label{35}
\end{equation}%
which are second-class and, essentially, irreducible. Following the line
exposed in \cite{EPL} it can be shown that the Dirac bracket associated with
the irreducible constraints (\ref{35}) takes the form
\begin{equation}
\left. \left[ F,G\right] ^{\left( 1\right) \ast }\right\vert _{\mathrm{ired}%
}=\left[ F,G\right] -\left[ F,\bar{\chi}_{\alpha _{0}}\right] \mu ^{\left(
1\right) \alpha _{0}\beta _{0}}\left[ \bar{\chi}_{\beta _{0}},G\right]
\label{31}
\end{equation}%
and it is (weakly) equal to the original Dirac bracket (\ref{10})
\begin{equation}
\left[ F,G\right] ^{\left( 1\right) \ast }\approx \left. \left[ F,G\right]
^{\left( 1\right) \ast }\right\vert _{\mathrm{ired}}.  \label{31zx}
\end{equation}%
In (\ref{31}) the quantities $\mu ^{\left( 1\right) \alpha _{0}\beta _{0}}$
are the elements of an invertible, antisymmetric matrix, expressed by
\begin{equation}
\mu ^{\left( 1\right) \alpha _{0}\beta _{0}}\approx M^{\left( 1\right)
\alpha _{0}\beta _{0}}+Z_{\lambda _{1}}^{\alpha _{0}}\bar{D}_{\beta
_{1}}^{\lambda _{1}}\Gamma ^{\beta _{1}\gamma _{1}}\bar{D}_{\gamma
_{1}}^{\sigma _{1}}Z_{\sigma _{1}}^{\beta _{0}},  \label{31qa}
\end{equation}%
with $\Gamma ^{\beta _{1}\gamma _{1}}$ the inverse of $\Gamma _{\alpha
_{1}\beta _{1}}$. Formula (\ref{31zx}) is essential in our context because
it proves that one can indeed approach first-order reducible second-class
constraints in an irreducible fashion.

In the case of second-order reducible second-class constraints, one
constructs the irreducible constraints%
\begin{equation}
\tilde{\chi}_{\alpha _{0}}=\chi _{\alpha _{0}}+A_{\alpha _{0}}^{\alpha
_{1}}Y_{\alpha _{1}}\approx 0,\qquad \tilde{\chi}_{\alpha _{2}}=Z_{\alpha
_{2}}^{\alpha _{1}}Y_{\alpha _{1}}\approx 0,  \label{28x}
\end{equation}%
where
\begin{equation}
A_{\sigma _{0}}^{\rho _{1}}=\hat{E}_{\alpha _{1}}^{\rho _{1}}\bar{A}_{\sigma
_{0}}^{\alpha _{1}},  \label{27y}
\end{equation}%
with $\hat{E}_{\alpha _{1}}^{\rho _{1}}$ the elements of an invertible
matrix \cite{JPA}. Following the line exposed in \cite{JPA} it can be shown
that the Dirac bracket associated with the irreducible constraints (\ref{28x}%
) takes the form%
\begin{eqnarray}
\left. \left[ F,G\right] ^{\left( 2\right) \ast }\right\vert _{\mathrm{ired}%
} &=&\left[ F,G\right] -\left[ F,\tilde{\chi}_{\alpha _{0}}\right] \mu
^{\left( 2\right) \alpha _{0}\beta _{0}}\left[ \tilde{\chi}_{\beta _{0}},G%
\right]  \notag \\
&&-\left[ F,\tilde{\chi}_{\alpha _{0}}\right] Z_{\gamma _{1}}^{\alpha _{0}}%
\hat{e}_{\sigma _{1}}^{\gamma _{1}}\Gamma ^{\sigma _{1}\lambda
_{1}}A_{\lambda _{1}}^{\tau _{2}}\bar{D}_{\tau _{2}}^{\beta _{2}}\left[
\tilde{\chi}_{\beta _{2}},G\right]  \notag \\
&&-\left[ F,\tilde{\chi}_{\alpha _{2}}\right] \bar{D}_{\lambda _{2}}^{\alpha
_{2}}A_{\sigma _{1}}^{\lambda _{2}}\Gamma ^{\sigma _{1}\lambda _{1}}\hat{e}%
_{\lambda _{1}}^{\gamma _{1}}Z_{\gamma _{1}}^{\beta _{0}}\left[ \tilde{\chi}%
_{\beta _{0}},G\right]  \notag \\
&&-\left[ F,\tilde{\chi}_{\alpha _{2}}\right] \bar{D}_{\lambda _{2}}^{\alpha
_{2}}A_{\sigma _{1}}^{\lambda _{2}}\Gamma ^{\sigma _{1}\lambda
_{1}}A_{\lambda _{1}}^{\tau _{2}}\bar{D}_{\tau _{2}}^{\beta _{2}}\left[
\tilde{\chi}_{\beta _{2}},G\right] ,  \label{i5}
\end{eqnarray}%
where
\begin{eqnarray}
&&\mu ^{\left( 2\right) \lambda _{0}\sigma _{0}}\approx M^{\left( 2\right)
\lambda _{0}\sigma _{0}}+Z_{\lambda _{1}}^{\lambda _{0}}\tilde{\omega}%
^{\lambda _{1}\sigma _{1}}Z_{\sigma _{1}}^{\sigma _{0}},  \label{21z} \\
&&\tilde{\omega}^{\alpha _{1}\beta _{1}}=\hat{e}_{\sigma _{1}}^{\alpha
_{1}}\Gamma ^{\sigma _{1}\tau _{1}}\hat{e}_{\tau _{1}}^{\beta _{1}}
\label{27w}
\end{eqnarray}%
and $\hat{e}_{\sigma _{1}}^{\alpha _{1}}$ are the elements of the inverse of
the matrix with the elements $\hat{E}_{\alpha _{1}}^{\gamma _{1}}$. In (\ref%
{i5}) the quantities denoted by $A_{\lambda _{1}}^{\tau _{2}}$ are some
functions chosen such that%
\begin{equation}
\mathrm{rank}\left( Z_{\alpha _{2}}^{\lambda _{1}}A_{\lambda _{1}}^{\tau
_{2}}\right) =M_{2}  \label{qew1}
\end{equation}%
and $\bar{D}_{\tau _{2}}^{\beta _{2}}$ stand for the elements of the inverse
of the matrix with the elements $Z_{\alpha _{2}}^{\lambda _{1}}A_{\lambda
_{1}}^{\tau _{2}}=D_{\alpha _{2}}^{\tau _{2}}$. Moreover, according to the
general proof from \cite{JPA}, one has%
\begin{equation}
\left[ F,G\right] ^{\left( 2\right) \ast }\approx \left. \left[ F,G\right]
^{\left( 2\right) \ast }\right\vert _{\mathrm{ired}},  \label{qwe2}
\end{equation}%
which shows that second-order reducible second-class constraints can also be
approached in an irreducible fashion.

\section{Third-order reducible second-class constraints\label{third}}

\subsection{Reducible approach\label{red3}}

\subsubsection{Dirac bracket for third-order reducible second-class
constraints\label{parDirac1}}

In this section we will consider third-order reducible second-class
constraints. This means that, beside the first-order reducibility relations (%
\ref{2}), the following relations also hold%
\begin{eqnarray}
Z_{\alpha _{2}}^{\alpha _{1}}Z_{\alpha _{1}}^{\alpha _{0}} &\approx
&0,\qquad \alpha _{2}=\overline{1,M_{2}},  \label{110x} \\
Z_{\alpha _{3}}^{\alpha _{2}}Z_{\alpha _{2}}^{\alpha _{1}} &\approx
&0,\qquad \alpha _{3}=\overline{1,M_{3}}.  \label{d2}
\end{eqnarray}%
They are known as the reducibility relations of order two and three,
respectively. In addition, all the third-order reducibility functions $%
Z_{\alpha _{3}}^{\alpha _{2}}$ are assumed to be independent. Under these
circumstances, the number of independent second-class constraint functions
is equal to $M\equiv M_{0}-M_{1}+M_{2}-M_{3}$. As a consequence, we can work
again with a Dirac bracket of the type (\ref{4}), but written in terms of $M$
independent functions $\chi _{A}$, i.e.
\begin{equation}
\left[ F,G\right] ^{\left( 3\right) \ast }=\left[ F,G\right] -\left[ F,\chi
_{A}\right] M^{\left( 3\right) AB}\left[ \chi _{B},G\right] ,\qquad A=%
\overline{1,M},  \label{110b}
\end{equation}%
where $C_{AB}^{\left( 3\right) }M^{\left( 3\right) BC}\approx \delta
_{A}^{C} $, with $C_{AB}^{\left( 3\right) }=\left[ \chi _{A},\chi _{B}\right]
$. It is clear that the matrix of elements
\begin{equation}
C_{\alpha _{0}\beta _{0}}^{\left( 3\right) }=\left[ \chi _{\alpha _{0}},\chi
_{\beta _{0}}\right]  \label{110d}
\end{equation}%
also satisfies the relations%
\begin{equation}
Z_{\alpha _{1}}^{\alpha _{0}}C_{\alpha _{0}\beta _{0}}^{\left( 3\right)
}\approx 0  \label{110e}
\end{equation}%
and, actually, its rank is equal to $M$.

Let $\bar{A}_{\alpha _{2}}^{\alpha _{3}}$ be a solution to
\begin{equation}
Z_{\alpha _{3}}^{\alpha _{2}}\bar{A}_{\alpha _{2}}^{\beta _{3}}\approx
\delta _{\alpha _{3}}^{\beta _{3}},  \label{a20}
\end{equation}%
and $\bar{\omega}_{\alpha _{2}\beta _{2}}=-\bar{\omega}_{\beta _{2}\alpha
_{2}}$ a solution to%
\begin{equation}
Z_{\alpha _{3}}^{\alpha _{2}}\bar{\omega}_{\alpha _{2}\beta _{2}}\approx 0.
\label{a10}
\end{equation}%
Then, we can introduce an antisymmetric matrix, of elements $\hat{\omega}%
^{\beta _{2}\gamma _{2}}$, defined through the relation%
\begin{equation}
\bar{\omega}_{\alpha _{2}\beta _{2}}\hat{\omega}^{\beta _{2}\gamma
_{2}}\approx \delta _{\alpha _{2}}^{\gamma _{2}}-\bar{A}_{\alpha
_{2}}^{\alpha _{3}}Z_{\alpha _{3}}^{\gamma _{2}}\equiv D_{\alpha
_{2}}^{\gamma _{2}}.  \label{a30}
\end{equation}%
If we take into account equation (\ref{a10}), then it can be checked that $%
\hat{\omega}^{\beta _{2}\gamma _{2}}$ are defined up to the transformation
\begin{equation}
\hat{\omega}^{\beta _{2}\gamma _{2}}\rightarrow \hat{\omega}^{\beta
_{2}\gamma _{2}}+Z_{\beta _{3}}^{\beta _{2}}\hat{q}^{\beta _{3}\gamma
_{3}}Z_{\gamma _{3}}^{\gamma _{2}},  \label{az11}
\end{equation}%
where $\hat{q}^{\beta _{3}\gamma _{3}}$ are some arbitrary, antisymmetric
functions. On the other hand, simple computation shows that the matrix of
elements $D_{\alpha _{2}}^{\gamma _{2}}$ satisfies the relations
\begin{eqnarray}
&&D_{\alpha _{2}}^{\gamma _{2}}\bar{A}_{\gamma _{2}}^{\gamma _{3}}\approx
0,\qquad Z_{\alpha _{3}}^{\alpha _{2}}D_{\alpha _{2}}^{\gamma _{2}}\approx 0,
\label{ax1} \\
&&D_{\alpha _{2}}^{\gamma _{2}}Z_{\gamma _{2}}^{\alpha _{1}}\approx
Z_{\alpha _{2}}^{\alpha _{1}},\qquad D_{\alpha _{2}}^{\gamma _{2}}D_{\gamma
_{2}}^{\beta _{2}}\approx D_{\alpha _{2}}^{\beta _{2}}.  \label{ay1}
\end{eqnarray}%
Based on the latter formula from (\ref{ax1}), we find that $D_{\alpha
_{2}}^{\gamma _{2}}$ can alternatively be expressed as%
\begin{equation}
D_{\alpha _{2}}^{\gamma _{2}}\approx Z_{\alpha _{2}}^{\alpha _{1}}\bar{A}%
_{\alpha _{1}}^{\gamma _{2}},  \label{10qa}
\end{equation}%
for some functions $\bar{A}_{\alpha _{1}}^{\gamma _{2}}$. We notice that the
above mentioned functions are defined up to the transformations%
\begin{equation}
\bar{A}_{\alpha _{1}}^{\alpha _{2}}\rightarrow \bar{A}_{\alpha _{1}}^{\alpha
_{2}}+\mu _{\alpha _{0}}^{\alpha _{2}}Z_{\alpha _{1}}^{\alpha _{0}},
\label{redefa12}
\end{equation}%
with $\mu _{\alpha _{0}}^{\gamma _{2}}$ some arbitrary functions.

Using now the former relation from (\ref{ay1}) and (\ref{10qa}), we deduce
that
\begin{equation}
Z_{\alpha _{2}}^{\alpha _{1}}D_{\alpha _{1}}^{\gamma _{1}}\approx 0,
\label{120b}
\end{equation}%
where
\begin{equation}
D_{\alpha _{1}}^{\gamma _{1}}\approx \delta _{\alpha _{1}}^{\gamma _{1}}-%
\bar{A}_{\alpha _{1}}^{\alpha _{2}}Z_{\alpha _{2}}^{\gamma _{1}}.
\label{110f}
\end{equation}%
Relations (\ref{120b}) and (\ref{110f}) ensure that $D_{\alpha _{1}}^{\gamma
_{1}}$\ is a `projection' (idempotent) in the weak sense%
\begin{equation}
D_{\beta _{1}}^{\alpha _{1}}D_{\gamma _{1}}^{\beta _{1}}\approx D_{\gamma
_{1}}^{\alpha _{1}}.  \label{210q}
\end{equation}%
With $D_{\alpha _{1}}^{\gamma _{1}}$ of the form (\ref{110f}) at hand, from (%
\ref{110x}) it follows that
\begin{equation}
D_{\alpha _{1}}^{\gamma _{1}}Z_{\gamma _{1}}^{\gamma _{0}}\approx Z_{\alpha
_{1}}^{\gamma _{0}}.  \label{110w}
\end{equation}%
Formula (\ref{120b}) emphasizes an alternative expression for $D_{\alpha
_{1}}^{\gamma _{1}}$%
\begin{equation}
D_{\alpha _{1}}^{\gamma _{1}}\approx Z_{\alpha _{1}}^{\alpha _{0}}\bar{A}%
_{\alpha _{0}}^{\gamma _{1}},  \label{1qaa}
\end{equation}%
for some functions $\bar{A}_{\alpha _{0}}^{\gamma _{1}}$. Accordingly, from (%
\ref{110w}) and (\ref{1qaa}) we find that
\begin{equation}
Z_{\alpha _{1}}^{\alpha _{0}}D_{\alpha _{0}}^{\gamma _{0}}\approx 0,
\label{12ba}
\end{equation}%
where
\begin{equation}
D_{\alpha _{0}}^{\gamma _{0}}\approx \delta _{\alpha _{0}}^{\gamma _{0}}-%
\bar{A}_{\alpha _{0}}^{\alpha _{1}}Z_{\alpha _{1}}^{\gamma _{0}}.
\label{11fa}
\end{equation}%
Just like before, from relations (\ref{12ba}) and (\ref{11fa}) we obtain
that $D_{\alpha _{0}}^{\gamma _{0}}$\ is also a `projection' in the weak
sense%
\begin{equation}
D_{\beta _{0}}^{\alpha _{0}}D_{\gamma _{0}}^{\beta _{0}}\approx D_{\gamma
_{0}}^{\alpha _{0}}.  \label{210qa}
\end{equation}

At this stage, we can rewrite the Dirac bracket expressed by (\ref{110b}) in
terms of all the second-class constraint functions. In view of this, we add
an antisymmetric matrix, of elements $M^{\left( 3\right) \alpha _{0}\beta
_{0}}$, through the relation
\begin{equation}
C_{\alpha _{0}\beta _{0}}^{\left( 3\right) }M^{\left( 3\right) \beta
_{0}\gamma _{0}}\approx D_{\alpha _{0}}^{\gamma _{0}},  \label{110c}
\end{equation}%
such that the formula
\begin{equation}
\left[ F,G\right] ^{\left( 3\right) \ast }=\left[ F,G\right] -\left[ F,\chi
_{\alpha _{0}}\right] M^{\left( 3\right) \alpha _{0}\beta _{0}}\left[ \chi
_{\beta _{0}},G\right]  \label{140q}
\end{equation}%
defines the same Dirac bracket like (\ref{110b}) on the surface (\ref{1}).
It is simple to see that the elements $M^{\left( 3\right) \alpha _{0}\beta
_{0}}$ are defined up to the transformation
\begin{equation}
M^{\left( 3\right) \alpha _{0}\beta _{0}}\rightarrow M^{\left( 3\right)
\alpha _{0}\beta _{0}}+Z_{\alpha _{1}}^{\alpha _{0}}p^{\alpha _{1}\beta
_{1}}Z_{\alpha _{1}}^{\beta _{0}},  \label{140r}
\end{equation}%
with $p^{\alpha _{1}\beta _{1}}$ some arbitrary, antisymmetric functions. We
notice that relations (\ref{110x}), (\ref{d2}), and (\ref{12ba}) ensure that%
\begin{equation}
\mathrm{rank}\left( D_{\gamma _{0}}^{\alpha _{0}}\right) \approx M
\label{120a}
\end{equation}%
and hence the rank of the matrix of elements $C_{\alpha _{0}\beta
_{0}}^{\left( 3\right) }M^{\left( 3\right) \beta _{0}\gamma _{0}}$ is equal
to the number of independent second-class constraints in the case of the
reducibility of order three. Meanwhile, we have that
\begin{equation}
\left[ \chi _{\alpha _{0}},G\right] ^{\left( 3\right) \ast }\approx -\bar{A}%
_{\alpha _{0}}^{\alpha _{1}}\left[ Z_{\alpha _{1}}^{\beta _{0}},G\right]
\chi _{\beta _{0}},  \label{q120}
\end{equation}%
so $\left[ \chi _{\alpha _{0}},G\right] ^{\left( 3\right) \ast }=0$, for any
$G$, on the surface of third-order reducible second-class constraints.

\subsubsection{Expressing the Dirac bracket in terms of an invertible matrix
\label{red3invert}}

Initially, we will establish some useful properties of the functions $\bar{A}%
_{\alpha _{0}}^{\alpha _{1}}$, $\bar{A}_{\alpha _{1}}^{\alpha _{2}}$, and $%
\bar{A}_{\alpha _{2}}^{\alpha _{3}}$. We introduce (\ref{10qa}) in the
former relation from (\ref{ax1}) and infer%
\begin{equation}
Z_{\alpha _{2}}^{\alpha _{1}}\bar{A}_{\alpha _{1}}^{\gamma _{2}}\bar{A}%
_{\gamma _{2}}^{\gamma _{3}}\approx 0,  \label{rnew1}
\end{equation}%
which implies the existence of some smooth functions $M_{\alpha
_{0}}^{\gamma _{3}}$\ such that%
\begin{equation}
\bar{A}_{\alpha _{1}}^{\gamma _{2}}\bar{A}_{\gamma _{2}}^{\gamma
_{3}}\approx M_{\alpha _{0}}^{\gamma _{3}}Z_{\alpha _{1}}^{\alpha _{0}}.
\label{rnew2}
\end{equation}%
On the other hand, the functions $\bar{A}_{\alpha _{1}}^{\alpha _{2}}$\
contain the ambiguities given in (\ref{redefa12}), which can be speculated
via choosing $\mu _{\alpha _{0}}^{\alpha _{2}}=-M_{\alpha _{0}}^{\gamma
_{3}}Z_{\gamma _{3}}^{\alpha _{2}}$\ such that these functions satisfy the
conditions%
\begin{equation}
\bar{A}_{\alpha _{1}}^{\alpha _{2}}\bar{A}_{\alpha _{2}}^{\alpha
_{3}}\approx 0.  \label{c16}
\end{equation}%
Using definition (\ref{a30}) and relations (\ref{110f}) and (\ref{c16}), we
obtain%
\begin{eqnarray}
\bar{A}_{\alpha _{1}}^{\alpha _{2}}D_{\alpha _{2}}^{\beta _{2}} &\approx &%
\bar{A}_{\alpha _{1}}^{\beta _{2}},  \label{c14} \\
D_{\alpha _{1}}^{\gamma _{1}}\bar{A}_{\gamma _{1}}^{\beta _{2}} &\approx &0.
\label{c15}
\end{eqnarray}%
By inserting now (\ref{c15}) in (\ref{1qaa}), we deduce the relation%
\begin{equation}
\bar{A}_{\alpha _{0}}^{\alpha _{1}}\bar{A}_{\alpha _{1}}^{\alpha
_{2}}\approx 0,  \label{waq12}
\end{equation}%
which enables us, by means of equations (\ref{110f}) and (\ref{11fa}), to
establish the formulas%
\begin{eqnarray}
D_{\tau _{0}}^{\alpha _{0}}\bar{A}_{\alpha _{0}}^{\alpha _{1}} &\approx &0,
\label{waq9} \\
\bar{A}_{\alpha _{0}}^{\alpha _{1}}D_{\alpha _{1}}^{\beta _{1}} &\approx &%
\bar{A}_{\alpha _{0}}^{\beta _{1}}.  \label{waq10}
\end{eqnarray}

Before expressing the Dirac bracket in terms of an invertible matrix, let us
analyze equations (\ref{a20}) and (\ref{a10}). The solution to (\ref{a20})
may be set under the form
\begin{equation}
\bar{A}_{\alpha _{2}}^{\alpha _{3}}\approx A_{\alpha _{2}}^{\beta _{3}}\bar{D%
}_{\beta _{3}}^{\alpha _{3}},  \label{a40}
\end{equation}%
where $A_{\alpha _{2}}^{\beta _{3}}$ are some functions taken such that the
matrix of elements
\begin{equation}
D_{\alpha _{3}}^{\gamma _{3}}=Z_{\alpha _{3}}^{\alpha _{2}}A_{\alpha
_{2}}^{\gamma _{3}}  \label{a50}
\end{equation}%
is of maximum rank%
\begin{equation}
\mathrm{rank}\left( D_{\alpha _{3}}^{\gamma _{3}}\right) =M_{3}.  \label{a60}
\end{equation}%
The notations $\bar{D}_{\beta _{3}}^{\alpha _{3}}$ stand for the elements of
the inverse of $D_{\alpha _{3}}^{\gamma _{3}}$\footnote{%
Strictly speaking, the solution to equation (\ref{a20}) has the general form
$\bar{A}_{\alpha _{2}}^{\alpha _{3}}\approx A_{\alpha _{2}}^{\lambda _{3}}%
\bar{D}_{\lambda _{3}}^{\alpha _{3}}+Z_{\alpha _{2}}^{\alpha _{1}}u_{\alpha
_{1}}^{\alpha _{3}}+\bar{\omega}_{\alpha _{2}\lambda _{2}}v^{\lambda
_{2}\alpha _{2}}$, where $u_{\alpha _{1}}^{\alpha _{3}}$ and $v^{\lambda
_{2}\alpha _{3}}$ are some arbitrary functions. If we make the redefinitions
$u_{\alpha _{1}}^{\alpha _{3}}=\hat{u}_{\alpha _{1}}^{\lambda _{3}}\bar{D}%
_{\lambda _{3}}^{\alpha _{3}}$ and$\;v^{\lambda _{2}\alpha _{3}}=\hat{v}%
^{\lambda _{2}\lambda _{3}}\bar{D}_{\lambda _{3}}^{\alpha _{3}}$, with $\hat{%
u}_{\alpha _{1}}^{\lambda _{3}}$ and $\hat{v}^{\lambda _{2}\lambda _{3}}$
some arbitrary functions, then we can bring $\bar{A}_{\alpha _{2}}^{\alpha
_{3}}$ to the form $\bar{A}_{\alpha _{2}}^{\alpha _{3}}\approx \left(
A_{\alpha _{2}}^{\lambda _{3}}+Z_{\alpha _{2}}^{\alpha _{1}}\hat{u}_{\alpha
_{1}}^{\lambda _{3}}+\bar{\omega}_{\alpha _{2}\lambda _{2}}\hat{v}^{\lambda
_{2}\lambda _{3}}\right) \bar{D}_{\lambda _{3}}^{\alpha _{3}}$. On the other
hand, the quantities $A_{\alpha _{2}}^{\lambda _{3}}$ taken such that the
rank of (\ref{a50}) is maximum are defined up to the transformation $%
A_{\alpha _{2}}^{\lambda _{3}}\rightarrow A_{\alpha _{2}}^{\prime \lambda
_{3}}=A_{\alpha _{2}}^{\lambda _{3}}+Z_{\alpha _{2}}^{\alpha _{1}}\tau
_{\alpha _{1}}^{\lambda _{3}}+\bar{\omega}_{\alpha _{2}\lambda _{2}}\lambda
^{\lambda _{2}\lambda _{3}}$, in the sense that $Z_{\beta _{3}}^{\alpha
_{2}}A_{\alpha _{2}}^{\lambda _{3}}\approx Z_{\beta _{3}}^{\alpha
_{2}}A_{\alpha _{2}}^{\prime \lambda _{3}}$, with $\tau _{\alpha
_{1}}^{\lambda _{3}}$ and $\lambda ^{\lambda _{2}\lambda _{3}}$ also
arbitrary. Thus, we can absorb the quantity $Z_{\alpha _{2}}^{\alpha _{1}}%
\hat{u}_{\alpha _{1}}^{\lambda _{3}}+\bar{\omega}_{\alpha _{2}\lambda _{2}}%
\hat{v}^{\lambda _{2}\lambda _{3}}$ from $\bar{A}_{\alpha _{2}}^{\alpha
_{3}} $ through a redefinition of $A_{\alpha _{2}}^{\lambda _{3}}$ and
finally obtain solution (\ref{a40}).}.

Using (\ref{a40}) in (\ref{a30}), we have
\begin{equation}
D_{\alpha _{2}}^{\gamma _{2}}\equiv \delta _{\alpha _{2}}^{\gamma
_{2}}-A_{\alpha _{2}}^{\beta _{3}}\bar{D}_{\beta _{3}}^{\alpha
_{3}}Z_{\alpha _{3}}^{\gamma _{2}},  \label{a70}
\end{equation}%
while (\ref{a40}) and the former relation from (\ref{ax1}) lead to
\begin{equation}
D_{\alpha _{2}}^{\gamma _{2}}A_{\gamma _{2}}^{\gamma _{3}}\approx 0.
\label{c20}
\end{equation}%
Inserting $D_{\alpha _{2}}^{\beta _{2}}$ given by (\ref{a70}) in (\ref{c14}%
), we deduce%
\begin{equation}
\bar{A}_{\gamma _{1}}^{\gamma _{2}}A_{\gamma _{2}}^{\gamma _{3}}\approx 0.
\label{c21}
\end{equation}%
Employing now the latter relation from (\ref{ax1}), we get that the solution
to equation (\ref{a10}) reads as
\begin{equation}
\bar{\omega}_{\alpha _{2}\beta _{2}}\approx D_{\alpha _{2}}^{\gamma _{2}}%
\tilde{\omega}_{\gamma _{2}\delta _{2}}D_{\beta _{2}}^{\delta _{2}},
\label{a82}
\end{equation}%
with $\tilde{\omega}_{\gamma _{2}\delta _{2}}$ the elements of an
antisymmetric matrix. Multiplying (\ref{a30}) with $A_{\gamma _{2}}^{\gamma
_{3}}$ and taking into account (\ref{c20}), we infer the equation%
\begin{equation}
\bar{\omega}_{\alpha _{2}\beta _{2}}\hat{\omega}^{\beta _{2}\gamma
_{2}}A_{\gamma _{2}}^{\gamma _{3}}\approx 0,  \label{a83}
\end{equation}%
whose solution is
\begin{equation}
\hat{\omega}^{\beta _{2}\gamma _{2}}A_{\gamma _{2}}^{\gamma _{3}}\approx
Z_{\beta _{3}}^{\beta _{2}}Q^{\beta _{3}\gamma _{3}}.  \label{q1}
\end{equation}%
Since the matrix of elements $\hat{\omega}^{\beta _{2}\gamma _{2}}$ is
defined up to transformation (\ref{az11}), we are free to make the choice $%
\hat{q}^{\beta _{3}\gamma _{3}}\approx -Q^{\beta _{3}\lambda _{3}}\bar{D}%
_{\lambda _{3}}^{\gamma _{3}}$, which brings the solution to equation (\ref%
{a83}) at the form
\begin{equation}
\hat{\omega}^{\beta _{2}\gamma _{2}}A_{\gamma _{2}}^{\gamma _{3}}\approx 0,
\label{q2}
\end{equation}%
which further implies
\begin{equation}
\hat{\omega}^{\alpha _{2}\beta _{2}}\approx D_{\rho _{2}}^{\alpha _{2}}%
\tilde{\omega}^{\rho _{2}\sigma _{2}}D_{\sigma _{2}}^{\beta _{2}},
\label{a84}
\end{equation}%
with $\tilde{\omega}^{\rho _{2}\sigma _{2}}$ the elements of an
antisymmetric matrix.

Under these conditions, the next theorem can be proved to hold.

\begin{theorem}
\label{th1} The matrices of elements $\tilde{\omega}_{\gamma _{2}\delta
_{2}} $ and $\tilde{\omega}^{\rho _{2}\sigma _{2}}$ can always be taken to
satisfy the following properties:

\noindent (a) invertibility;

\noindent (b) fulfillment of relation
\begin{equation}
\tilde{\omega}^{\rho _{2}\sigma _{2}}D_{\sigma _{2}}^{\gamma _{2}}\tilde{%
\omega}_{\gamma _{2}\delta _{2}}\approx D_{\delta _{2}}^{\rho _{2}}.
\label{a85}
\end{equation}
\end{theorem}

\textbf{Proof. }(a) Inserting the latter relation from (\ref{ay1}) in (\ref%
{a82}) and (\ref{a84}), we reach the equations%
\begin{eqnarray}
D_{\alpha _{2}}^{\gamma _{2}}\bar{\omega}_{\gamma _{2}\delta _{2}}D_{\beta
_{2}}^{\delta _{2}} &\approx &D_{\alpha _{2}}^{\gamma _{2}}\tilde{\omega}%
_{\gamma _{2}\delta _{2}}D_{\beta _{2}}^{\delta _{2}},  \label{a190} \\
D_{\rho _{2}}^{\alpha _{2}}\hat{\omega}^{\rho _{2}\sigma _{2}}D_{\sigma
_{2}}^{\beta _{2}} &\approx &D_{\rho _{2}}^{\alpha _{2}}\tilde{\omega}^{\rho
_{2}\sigma _{2}}D_{\sigma _{2}}^{\beta _{2}},  \label{a191}
\end{eqnarray}%
which give
\begin{eqnarray}
\tilde{\omega}_{\gamma _{2}\delta _{2}} &\approx &\bar{\omega}_{\gamma
_{2}\delta _{2}}+A_{\gamma _{2}}^{\gamma _{3}}\bar{D}_{\gamma _{3}}^{\rho
_{3}}\xi _{\rho _{3}\sigma _{3}}\bar{D}_{\delta _{3}}^{\sigma _{3}}A_{\delta
_{2}}^{\delta _{3}},  \label{a121} \\
\tilde{\omega}^{\rho _{2}\sigma _{2}} &\approx &\hat{\omega}^{\rho
_{2}\sigma _{2}}+Z_{\rho _{3}}^{\rho _{2}}\xi ^{\rho _{3}\sigma
_{3}}Z_{\sigma _{3}}^{\sigma _{2}},  \label{a122}
\end{eqnarray}%
with $\xi _{\rho _{3}\sigma _{3}}$ and $\xi ^{\rho _{3}\sigma _{3}}$ the
elements of some invertible, antisymmetric matrices. With the help of (\ref%
{a10}), (\ref{a30}), and (\ref{q2}) and relying on relations (\ref{a121})
and (\ref{a122}) we find%
\begin{equation}
\tilde{\omega}_{\gamma _{2}\delta _{2}}\tilde{\omega}^{\delta _{2}\sigma
_{2}}\approx D_{\gamma _{2}}^{\sigma _{2}}+Z_{\gamma _{2}}^{\gamma _{3}}\bar{%
D}_{\gamma _{3}}^{\rho _{3}}A_{\rho _{3}}^{\sigma _{2}}.  \label{c27}
\end{equation}%
As $D_{\gamma _{2}}^{\sigma _{2}}$ is of the form (\ref{a70}), we find
immediately%
\begin{equation}
\tilde{\omega}_{\gamma _{2}\delta _{2}}\tilde{\omega}^{\delta _{2}\sigma
_{2}}\approx \delta _{\gamma _{2}}^{\sigma _{2}},  \label{c28}
\end{equation}%
which proves (a).

(b) Simple computation outputs
\begin{eqnarray}
&\tilde{\omega}^{\rho _{2}\sigma _{2}}D_{\sigma _{2}}^{\beta _{2}} \approx
\hat{\omega}^{\rho _{2}\beta _{2}},  \label{m5} \\
&\hat{\omega}^{\rho _{2}\beta _{2}}\tilde{\omega}_{\beta _{2}\lambda _{2}}
\approx \hat{\omega}^{\rho _{2}\beta _{2}}\bar{\omega}_{\beta _{2}\lambda
_{2}}\approx D_{\lambda _{2}}^{\rho _{2}},  \label{m6}
\end{eqnarray}%
which further imply
\begin{equation}
\tilde{\omega}^{\rho _{2}\sigma _{2}}D_{\sigma _{2}}^{\beta _{2}}\tilde{%
\omega}_{\beta _{2}\lambda _{2}}\approx D_{\lambda _{2}}^{\rho _{2}},
\label{m7}
\end{equation}%
such that (b) is also proved.$\Box $

Let $\bar{\omega}_{\alpha _{1}\beta _{1}}=-\bar{\omega}_{\beta _{1}\alpha
_{1}}$ be a solution to the equation
\begin{equation}
Z_{\alpha _{2}}^{\alpha _{1}}\bar{\omega}_{\alpha _{1}\beta _{1}}\approx 0.
\label{a1a}
\end{equation}%
Then, one can introduce an antisymmetric matrix, of elements $\hat{\omega}%
^{\beta _{1}\gamma _{1}}$, through the relation%
\begin{equation}
\bar{\omega}_{\alpha _{1}\beta _{1}}\hat{\omega}^{\beta _{1}\gamma
_{1}}\approx D_{\alpha _{1}}^{\gamma _{1}}.  \label{a3a}
\end{equation}%
Due to (\ref{a1a}), we conclude that the elements $\hat{\omega}^{\beta
_{1}\gamma _{1}}$ are defined up to the transformation
\begin{equation}
\hat{\omega}^{\beta _{1}\gamma _{1}}\rightarrow \hat{\omega}^{\beta
_{1}\gamma _{1}}+Z_{\beta _{2}}^{\beta _{1}}\hat{q}^{\beta _{2}\gamma
_{2}}Z_{\gamma _{2}}^{\gamma _{1}},  \label{az1}
\end{equation}%
with $\hat{q}^{\beta _{2}\gamma _{2}}$ some antisymmetric, but otherwise
arbitrary functions. Recalling relation (\ref{120b}), we obtain that the
solution to (\ref{a1a}) can be expressed as
\begin{equation}
\bar{\omega}_{\alpha _{1}\beta _{1}}\approx D_{\alpha _{1}}^{\gamma _{1}}%
\tilde{\omega}_{\gamma _{1}\delta _{1}}D_{\beta _{1}}^{\delta _{1}},
\label{a150}
\end{equation}%
with $\tilde{\omega}_{\gamma _{1}\delta _{1}}$ the elements of an
antisymmetric matrix. Acting with $\bar{A}_{\gamma _{1}}^{\gamma _{2}}$ on (%
\ref{a3a}) and taking into account the result given by (\ref{c15}), we infer
the equation%
\begin{equation}
\bar{\omega}_{\alpha _{1}\beta _{1}}\hat{\omega}^{\beta _{1}\gamma _{1}}\bar{%
A}_{\gamma _{1}}^{\gamma _{2}}\approx 0,  \label{a160}
\end{equation}%
whose solution reads as
\begin{equation}
\hat{\omega}^{\beta _{1}\gamma _{1}}\bar{A}_{\gamma _{1}}^{\gamma
_{2}}\approx Z_{\beta _{2}}^{\beta _{1}}Q^{\beta _{2}\gamma _{2}}.
\label{q3}
\end{equation}%
Due to the fact that the matrix of elements $\hat{\omega}^{\beta _{1}\gamma
_{1}}$ is defined up to transformation (\ref{az1}), we are free to make the
choice $\hat{q}^{\beta _{2}\gamma _{2}}\approx -Q^{\beta _{2}\gamma _{2}}$,
which brings equation (\ref{a160}) at the form
\begin{equation}
\hat{\omega}^{\beta _{1}\gamma _{1}}\bar{A}_{\gamma _{1}}^{\gamma
_{2}}\approx 0,  \label{q4}
\end{equation}%
such that its solution can be taken as\footnote{%
In fact, the general solution to equation (\ref{a160}) has the expression $%
\hat{\omega}^{\alpha _{1}\beta _{1}}=D_{\rho _{1}}^{\alpha _{1}}\tilde{\omega%
}^{\rho _{1}\sigma _{1}}D_{\sigma _{1}}^{\beta _{1}}+Z_{\alpha _{2}}^{\alpha
_{1}}u^{\alpha _{2}\beta _{2}}Z_{\beta _{2}}^{\beta _{1}}$, for some
antisymmetric functions $u^{\alpha _{2}\beta _{2}}$. But the quantities $%
\hat{\omega}^{\alpha _{1}\beta _{1}}$ are defined up to transformation (\ref%
{az1}), so one can absorb the terms $Z_{\alpha _{2}}^{\alpha _{1}}u^{\alpha
_{2}\beta _{2}}Z_{\beta _{2}}^{\beta _{1}}$ through a redefinition of $\hat{%
\omega}^{\alpha _{1}\beta _{1}}$ and obtain in the end precisely solution (%
\ref{a170}).}
\begin{equation}
\hat{\omega}^{\beta _{1}\gamma _{1}}=D_{\lambda _{1}}^{\beta _{1}}\tilde{%
\omega}^{\lambda _{1}\rho _{1}}D_{\rho _{1}}^{\gamma _{1}},  \label{a170}
\end{equation}%
with $\tilde{\omega}^{\lambda _{1}\rho _{1}}$ the elements of an
antisymmetric matrix.

Except from being antisymmetric, the matrices of elements $\tilde{\omega}%
_{\gamma _{1}\delta _{1}}$ and respectively $\tilde{\omega}^{\lambda
_{1}\rho _{1}}$ are arbitrary at this stage. The next theorem shows that
they are in fact related.

\begin{theorem}
\label{th2} The matrices of elements $\tilde{\omega}_{\gamma _{1}\delta
_{1}} $ and $\tilde{\omega}^{\lambda _{1}\rho _{1}}$ can always be taken to
satisfy the following properties:

\noindent (a) invertibility;

\noindent (b) fulfillment of relation
\begin{equation}
\tilde{\omega}^{\lambda _{1}\rho _{1}}D_{\rho _{1}}^{\gamma _{1}}\tilde{%
\omega}_{\gamma _{1}\delta _{1}}\approx D_{\delta _{1}}^{\lambda _{1}}.
\label{a180}
\end{equation}
\end{theorem}

\textbf{Proof. }(a) Substituting the latter relation from (\ref{210q}) in (%
\ref{a150}) and (\ref{a170}), we obtain the equations%
\begin{eqnarray}
D_{\alpha _{1}}^{\gamma _{1}}\bar{\omega}_{\gamma _{1}\delta _{1}}D_{\beta
_{1}}^{\delta _{1}} &\approx &D_{\alpha _{1}}^{\gamma _{1}}\tilde{\omega}%
_{\gamma _{1}\delta _{1}}D_{\beta _{1}}^{\delta _{1}},  \label{a1900} \\
D_{\rho _{1}}^{\alpha _{1}}\hat{\omega}^{\rho _{1}\sigma _{1}}D_{\sigma
_{1}}^{\beta _{1}} &\approx &D_{\rho _{1}}^{\alpha _{1}}\tilde{\omega}^{\rho
_{1}\sigma _{1}}D_{\sigma _{1}}^{\beta _{1}},  \label{a200}
\end{eqnarray}%
which then give
\begin{eqnarray}
\tilde{\omega}_{\gamma _{1}\delta _{1}} &\approx &\bar{\omega}_{\gamma
_{1}\delta _{1}}+\bar{A}_{\gamma _{1}}^{\gamma _{2}}\xi _{\gamma _{2}\delta
_{2}}\bar{A}_{\delta _{1}}^{\delta _{2}},  \label{a210} \\
\tilde{\omega}^{\rho _{1}\sigma _{1}} &\approx &\hat{\omega}^{\rho
_{1}\sigma _{1}}+Z_{\rho _{2}}^{\rho _{1}}\xi ^{\rho _{2}\sigma
_{2}}Z_{\sigma _{2}}^{\sigma _{1}},  \label{a220}
\end{eqnarray}%
with $\xi _{\gamma _{2}\delta _{2}}$ and $\xi ^{\rho _{2}\sigma _{2}}$ the
elements of some antisymmetric matrices, taken to be invertible. Each of the
terms from the right-hand sides of relations (\ref{a210}) and (\ref{a220})
possesses null vectors. It is known that the null vectors of $\bar{\omega}%
_{\gamma _{1}\delta _{1}}$ and $\hat{\omega}^{\rho _{1}\sigma _{1}}$ are $%
Z_{\alpha _{2}}^{\gamma _{1}}$ and $\bar{A}_{\sigma _{1}}^{\beta _{2}}$
respectively (see (\ref{a1a}) and (\ref{q4})), while $\bar{A}_{\gamma
_{1}}^{\gamma _{2}}\xi _{\gamma _{2}\delta _{2}}\bar{A}_{\delta
_{1}}^{\delta _{2}}$ and $Z_{\rho _{2}}^{\rho _{1}}\xi ^{\rho _{2}\sigma
_{2}}Z_{\sigma _{2}}^{\sigma _{1}}$ display the null vectors $\bar{A}%
_{\gamma _{0}}^{\gamma _{1}}$ and $Z_{\sigma _{1}}^{\sigma _{0}}$
respectively\footnote{%
The most general form of the null vectors corresponding to $\bar{\omega}
_{\gamma _{1}\delta _{1}}$ and $\hat{\omega}^{\rho _{1}\sigma _{1}}$ is of
the type $\nu ^{\gamma _{2}}Z_{\gamma _{2}}^{\gamma _{1}}$ and $A_{\sigma
_{1}}^{\sigma _{2}}\xi _{\sigma _{2}}$ respectively, with $\nu ^{\gamma
_{2}} $ and $\xi _{\sigma _{2}}$ arbitrary functions. Along the same line,
the functions $\bar{A}_{\gamma _{1}}^{\gamma _{2}}\xi _{\gamma _{2}\delta
_{2}}\bar{A}_{\delta _{1}}^{\delta _{2}}$ and $Z_{\rho _{2}}^{\rho _{1}}\xi
^{\rho _{2}\sigma _{2}}Z_{\sigma _{2}}^{\sigma _{1}}\ $ display (the most
general) null vectors $\tau ^{\gamma _{0}}\bar{A}_{\gamma _{0}}^{\gamma
_{1}} $ and $Z_{\sigma _{1}}^{\sigma _{0}}\kappa _{\sigma _{0}}$
respectively, with $\tau ^{\gamma _{0}}$ and $\kappa _{\sigma _{0}}$
arbitrary functions. However, these observations do not affect the proof in
any way.}. For this reason, the only candidates for null vectors of $\tilde{%
\omega}_{\gamma _{1}\delta _{1}}$ and $\tilde{\omega}^{\rho _{1}\sigma _{1}}$
are on the one hand $Z_{\alpha _{2}}^{\gamma _{1}}$ and $\bar{A}_{\sigma
_{1}}^{\beta _{2}}$ respectively and on the other hand $\bar{A}_{\gamma
_{0}}^{\gamma _{1}}$ and $Z_{\sigma _{1}}^{\sigma _{0}}$ respectively. We
show that none of these candidates are null vectors.

Indeed, from (\ref{a210}) and (\ref{a220}) we find%
\begin{eqnarray}
Z_{\alpha _{2}}^{\gamma _{1}}\tilde{\omega}_{\gamma _{1}\delta _{1}}
&\approx &D_{\alpha _{2}}^{\gamma _{2}}\xi _{\gamma _{2}\delta _{2}}\bar{A}%
_{\delta _{1}}^{\delta _{2}},  \label{a230} \\
\tilde{\omega}^{\rho _{1}\sigma _{1}}\bar{A}_{\sigma _{1}}^{\beta _{2}}
&\approx &Z_{\rho _{2}}^{\rho _{1}}\xi ^{\rho _{2}\sigma _{2}}D_{\sigma
_{2}}^{\beta _{2}}.  \label{a240}
\end{eqnarray}%
The right-hand sides of (\ref{a230}) and (\ref{a240}) are (weakly) vanishing
for
\begin{eqnarray}
\xi _{\gamma _{2}\delta _{2}} &=&A_{\gamma _{2}}^{\gamma _{3}}\theta
_{\gamma _{3}\delta _{3}}A_{\delta _{2}}^{\delta _{3}},  \label{a250} \\
\xi ^{\rho _{2}\sigma _{2}} &=&Z_{\rho _{3}}^{\rho _{2}}\theta ^{\rho
_{3}\sigma _{3}}Z_{\sigma _{3}}^{\sigma _{2}},  \label{a260}
\end{eqnarray}%
with $\theta _{\gamma _{3}\delta _{3}}$ and $\theta ^{\rho _{3}\sigma _{3}}$
the elements of some antisymmetric matrices. It is simple to see that the
matrices of elements $\xi _{\gamma _{2}\delta _{2}}$ and $\xi ^{\rho
_{2}\sigma _{2}}$ given by (\ref{a250}) and (\ref{a260}) respectively are
degenerate\footnote{%
The matrix of elements $\xi _{\gamma _{2}\delta _{2}}$ displays the null
vectors $u^{\gamma _{1}}\bar{A}_{\gamma _{1}}^{\gamma _{2}}$ and that of
elements $\xi ^{\rho _{2}\sigma _{2}}$ exhibits the null vectors $v_{\rho
_{1}}Z_{\rho _{2}}^{\rho _{1}}$.}, which contradicts the hypothesis on their
invertibility. Thus, it follows that the matrices of elements $\xi _{\gamma
_{2}\delta _{2}}$ and $\xi ^{\rho _{2}\sigma _{2}}$ cannot be expressed as
in (\ref{a250}) and (\ref{a260}) respectively. In consequence, the
quantities $Z_{\alpha _{2}}^{\gamma _{1}}\tilde{\omega}_{\gamma _{1}\delta
_{1}}$ and $\tilde{\omega}^{\rho _{1}\sigma _{1}}\bar{A}_{\sigma
_{1}}^{\beta _{2}}$ given in (\ref{a230}) and (\ref{a240}) respectively
cannot vanish, so the matrices of elements $\tilde{\omega}_{\gamma
_{1}\delta _{1}}$ and $\tilde{\omega}^{\rho _{1}\sigma _{1}}$ do not have
the functions $Z_{\alpha _{2}}^{\gamma _{1}}$ and $\bar{A}_{\sigma
_{1}}^{\beta _{2}}$ as null vectors respectively. Multiplying (\ref{a210})
by $\bar{A}_{\gamma _{0}}^{\gamma _{1}}$ and (\ref{a220}) by $Z_{\sigma
_{1}}^{\sigma _{0}}$, we deduce that%
\begin{eqnarray}
\bar{A}_{\gamma _{0}}^{\gamma _{1}}\tilde{\omega}_{\gamma _{1}\delta _{1}}
&\approx &\bar{A}_{\gamma _{0}}^{\gamma _{1}}\bar{\omega}_{\gamma _{1}\delta
_{1}},  \label{b250} \\
\tilde{\omega}^{\rho _{1}\sigma _{1}}Z_{\sigma _{1}}^{\sigma _{0}} &\approx &%
\hat{\omega}^{\rho _{1}\sigma _{1}}Z_{\sigma _{1}}^{\sigma _{0}}.
\label{b260}
\end{eqnarray}%
The right-hand sides of (\ref{b250}) and (\ref{b260}) vanish for
\begin{eqnarray}
\bar{\omega}_{\gamma _{1}\delta _{1}} &=&\bar{A}_{\gamma _{1}}^{\gamma _{2}}%
\bar{\theta}_{\gamma _{2}\delta _{2}}\bar{A}_{\delta _{1}}^{\delta _{2}},
\label{c25} \\
\hat{\omega}^{\rho _{1}\sigma _{1}} &=&Z_{\rho _{2}}^{\rho _{1}}\hat{\theta}%
^{\rho _{2}\sigma _{2}}Z_{\sigma _{2}}^{\sigma _{1}},  \label{c26}
\end{eqnarray}%
with $\bar{\theta}_{\gamma _{2}\delta _{2}}$ and $\hat{\theta}^{\rho
_{2}\sigma 2}$ the elements of some antisymmetric matrices. It is now easy
to see that neither $\bar{\omega}_{\gamma _{1}\delta _{1}}$ nor $\hat{\omega}%
^{\rho _{1}\sigma _{1}}$, given by (\ref{c25}) and (\ref{c26}) respectively,
can be brought to the form expressed by relations (\ref{a150}) and (\ref%
{a170}) respectively, for any choice of $\bar{\theta}_{\gamma _{2}\delta
_{2}}$ or $\hat{\theta}^{\rho _{2}\sigma _{2}}$. Thus, it follows that
neither of relations (\ref{c25}) or (\ref{c26}) can hold, so neither of the
quantities $\bar{A}_{\gamma _{0}}^{\gamma _{1}}\bar{\omega}_{\gamma
_{1}\delta _{1}}$ or $\hat{\omega}^{\rho _{1}\sigma _{1}}Z_{\sigma
_{1}}^{\sigma _{0}}$ can vanish. This further implies that the matrices of
elements $\tilde{\omega}_{\gamma _{1}\delta _{1}}$ and $\tilde{\omega}^{\rho
_{1}\sigma _{1}}$ do not possess the functions $\bar{A}_{\gamma
_{0}}^{\gamma _{1}}$ and $Z_{\sigma _{1}}^{\sigma _{0}}$ as null vectors
respectively, so we conclude that both the matrices of elements $\tilde{%
\omega}_{\gamma _{1}\delta _{1}}$ and $\tilde{\omega}^{\rho _{1}\sigma _{1}}$
(having the expressions (\ref{a210}) and (\ref{a220}) respectively) are
invertible.

Because of results (\ref{a1a}), (\ref{a3a}), and (\ref{q4}), from relations (%
\ref{a210}) and (\ref{a220}) one gets
\begin{equation}
\tilde{\omega}_{\gamma _{1}\delta _{1}}\tilde{\omega}^{\delta _{1}\sigma
_{1}}\approx D_{\gamma _{1}}^{\sigma _{1}}+\bar{A}_{\gamma _{1}}^{\gamma
_{2}}\xi _{\gamma _{2}\rho _{2}}D_{\lambda _{2}}^{\rho _{2}}\xi ^{\lambda
_{2}\sigma _{2}}Z_{\sigma _{2}}^{\sigma _{1}}.  \label{c66}
\end{equation}%
We take the functions $\xi _{\gamma _{2}\rho _{2}}$ and $\xi ^{\lambda
_{2}\sigma _{2}}$ of the form%
\begin{equation}
\xi _{\gamma _{2}\rho _{2}}=\tilde{\omega}_{\gamma _{2}\rho _{2}},\qquad \xi
^{\lambda _{2}\sigma _{2}}=\tilde{\omega}^{\lambda _{2}\sigma _{2}},
\label{c67}
\end{equation}%
which replaced in (\ref{c66}) leads (also due to (\ref{a85})) to
\begin{equation}
\tilde{\omega}_{\gamma _{1}\delta _{1}}\tilde{\omega}^{\delta _{1}\sigma
_{1}}\approx \delta _{\gamma _{1}}^{\sigma _{1}}.  \label{c68}
\end{equation}%
This proves (a).

(b) By straightforward computation, it results
\begin{eqnarray}
&\tilde{\omega}^{\rho _{1}\sigma _{1}}D_{\sigma _{1}}^{\lambda _{1}} \approx
\hat{\omega}^{\rho _{1}\lambda _{1}},  \label{a270} \\
&\hat{\omega}^{\rho _{1}\lambda _{1}}\tilde{\omega}_{\lambda _{1}\delta
_{1}} \approx \hat{\omega}^{\rho _{1}\lambda _{1}}\bar{\omega}_{\lambda
_{1}\delta _{1}}\approx D_{\delta _{1}}^{\rho _{1}},  \label{a280}
\end{eqnarray}%
which further yields
\begin{equation}
\tilde{\omega}^{\rho _{1}\sigma _{1}}D_{\sigma _{1}}^{\lambda _{1}}\tilde{%
\omega}_{\lambda _{1}\delta _{1}}\approx D_{\delta _{1}}^{\rho _{1}}
\label{a290}
\end{equation}%
and proves (b).$\Box $

With these elements at hand, the next theorem is shown to hold.

\begin{theorem}
\label{th3} There exists an invertible, antisymmetric matrix of elements $%
\mu ^{\left( 3\right) \alpha _{0}\beta _{0}}$ such that Dirac bracket (\ref%
{140q}) takes the form
\begin{equation}
\left[ F,G\right] ^{\left( 3\right) \ast }=\left[ F,G\right] -\left[ F,\chi
_{\alpha _{0}}\right] \mu ^{\left( 3\right) \alpha _{0}\beta _{0}}\left[
\chi _{\beta _{0}},G\right]  \label{240}
\end{equation}%
on the surface (\ref{1}).
\end{theorem}

\textbf{Proof. }First, we observe that $D_{\gamma _{0}}^{\alpha _{0}}$ given
in (\ref{11fa}) satisfies the relations
\begin{equation}
D_{\gamma _{0}}^{\alpha _{0}}\chi _{\alpha _{0}}\approx \chi _{\gamma _{0}}.
\label{170}
\end{equation}%
Multiplying (\ref{110c}) by $\bar{A}_{\gamma _{0}}^{\gamma _{1}}$ and using (%
\ref{waq9}), we obtain the equation
\begin{equation}
C_{\alpha _{0}\beta _{0}}^{\left( 3\right) }M^{\left( 3\right) \beta
_{0}\gamma _{0}}\bar{A}_{\gamma _{0}}^{\gamma _{1}}\approx 0,  \label{u1}
\end{equation}%
which then leads to
\begin{equation}
M^{\left( 3\right) \beta _{0}\gamma _{0}}\bar{A}_{\gamma _{0}}^{\gamma
_{1}}\approx Z_{\beta _{1}}^{\beta _{0}}f^{\beta _{1}\gamma _{1}},
\label{190}
\end{equation}%
for some functions $f^{\beta _{1}\gamma _{1}}$. Acting with $D_{\beta
_{0}}^{\tau _{0}}$ on (\ref{190}) and employing (\ref{12ba}), we find the
relation
\begin{equation}
M^{\left( 3\right) \beta _{0}\gamma _{0}}\bar{A}_{\gamma _{0}}^{\gamma
_{1}}D_{\beta _{0}}^{\tau _{0}}\approx 0,  \label{190x}
\end{equation}%
with the help of which (via formula (\ref{waq9})) we can write
\begin{equation}
M^{\left( 3\right) \beta _{0}\gamma _{0}}D_{\beta _{0}}^{\tau _{0}}\approx
D_{\beta _{0}}^{\gamma _{0}}\lambda ^{\beta _{0}\tau _{0}},  \label{190y}
\end{equation}%
for some $\lambda ^{\beta _{0}\tau _{0}}$. Acting now with $D_{\beta
_{0}}^{\tau _{0}}$ on (\ref{110c}) and taking into account (\ref{190y}), we
deduce
\begin{equation}
-C_{\alpha _{0}\gamma _{0}}^{\left( 3\right) }D_{\beta _{0}}^{\gamma
_{0}}\lambda ^{\beta _{0}\tau _{0}}\approx D_{\alpha _{0}}^{\tau _{0}}.
\label{190w}
\end{equation}%
On the other hand, relation (\ref{170}) implies
\begin{equation}
D_{\alpha _{0}}^{\beta _{0}}C_{\beta _{0}\gamma _{0}}^{\left( 3\right)
}\approx C_{\alpha _{0}\gamma _{0}}^{\left( 3\right) },  \label{190q}
\end{equation}%
such that, on behalf of (\ref{190w}) and (\ref{190q}), we have
\begin{equation}
-C_{\alpha _{0}\beta _{0}}^{\left( 3\right) }\lambda ^{\beta _{0}\tau
_{0}}\approx D_{\alpha _{0}}^{\tau _{0}}.  \label{190z}
\end{equation}%
Comparing (\ref{190z}) with (\ref{110c}) and using the fact that the
functions $M^{\left( 3\right) \alpha _{0}\beta _{0}}$ are defined up to
transformation (\ref{140r}), we infer the relation%
\begin{equation}
M^{\left( 3\right) \beta _{0}\tau _{0}}=-\lambda ^{\beta _{0}\tau _{0}},
\label{190u}
\end{equation}%
which substituted in (\ref{190y}) provides the equation%
\begin{equation}
M^{\left( 3\right) \beta _{0}\gamma _{0}}D_{\beta _{0}}^{\tau _{0}}\approx
M^{\left( 3\right) \tau _{0}\beta _{0}}D_{\beta _{0}}^{\gamma _{0}}.
\label{190t}
\end{equation}%
Using one more time the fact that the elements $M^{\left( 3\right) \alpha
_{0}\beta _{0}}$ are defined up to (\ref{140r}), from (\ref{190t}) we get
\begin{equation}
M^{\left( 3\right) \alpha _{0}\beta _{0}}\approx D_{\lambda _{0}}^{\alpha
_{0}}\mu ^{\left( 3\right) \lambda _{0}\sigma _{0}}D_{\sigma _{0}}^{\beta
_{0}},  \label{200}
\end{equation}%
where $\mu ^{\left( 3\right) \lambda _{0}\sigma _{0}}$ is an antisymmetric
matrix. Due to formula (\ref{waq9}) and relation (\ref{200}) we can write
\begin{equation}
M^{\left( 3\right) \alpha _{0}\beta _{0}}\bar{A}_{\beta _{0}}^{\gamma
_{1}}\approx 0.  \label{200y}
\end{equation}%
Inserting the former relation from (\ref{210q}) in (\ref{200}), we deduce%
\begin{equation}
D_{\lambda _{0}}^{\alpha _{0}}M^{\left( 3\right) \lambda _{0}\sigma
_{0}}D_{\sigma _{0}}^{\beta _{0}}\approx D_{\lambda _{0}}^{\alpha _{0}}\mu
^{\left( 3\right) \lambda _{0}\sigma _{0}}D_{\sigma _{0}}^{\beta _{0}},
\label{200x}
\end{equation}%
which further yields
\begin{equation}
\mu ^{\left( 3\right) \lambda _{0}\sigma _{0}}\approx M^{\left( 3\right)
\lambda _{0}\sigma _{0}}+Z_{\lambda _{1}}^{\lambda _{0}}\nu ^{\lambda
_{1}\sigma _{1}}Z_{\sigma _{1}}^{\sigma _{0}},  \label{210}
\end{equation}%
for an antisymmetric matrix, of elements $\nu ^{\lambda _{1}\sigma _{1}}$.
Now, we show that the matrix of elements $\mu ^{\left( 3\right) \lambda
_{0}\sigma _{0}}$ can be taken to be invertible. If we take $\nu ^{\lambda
_{1}\sigma _{1}}$ under the form $\nu ^{\lambda _{1}\sigma _{1}}=\tilde{%
\omega}^{\lambda _{1}\sigma _{1}}$, where $\tilde{\omega}^{\lambda
_{1}\sigma _{1}}$ are precisely the elements of the invertible matrix given
in (\ref{a220}), then we find directly
\begin{equation}
\mu ^{\left( 3\right) \lambda _{0}\sigma _{0}}\approx M^{\left( 3\right)
\lambda _{0}\sigma _{0}}+Z_{\lambda _{1}}^{\lambda _{0}}\tilde{\omega}%
^{\lambda _{1}\sigma _{1}}Z_{\sigma _{1}}^{\sigma _{0}}.  \label{210z}
\end{equation}%
Next, we show that the matrix of elements
\begin{equation}
\mu _{\rho _{0}\lambda _{0}}^{\left( 3\right) }\approx C_{\rho _{0}\lambda
_{0}}^{\left( 3\right) }+\bar{A}_{\rho _{0}}^{\rho _{1}}\tilde{\omega}_{\rho
_{1}\tau _{1}}\bar{A}_{\lambda _{0}}^{\tau _{1}},  \label{210x}
\end{equation}%
where $\tilde{\omega}_{\rho _{1}\tau _{1}}$ determines the invertible matrix
given in (\ref{a210}), is nothing but the inverse of the matrix of elements $%
\mu ^{\left( 3\right) \lambda _{0}\sigma _{0}}$ expressed by (\ref{210z}).
Indeed, from (\ref{110e}), (\ref{1qaa}), (\ref{110c}), and (\ref{200y}),
direct computation provides
\begin{equation}
\mu _{\rho _{0}\lambda _{0}}^{\left( 3\right) }\mu ^{\left( 3\right) \lambda
_{0}\sigma _{0}}\approx D_{\rho _{0}}^{\sigma _{0}}+\bar{A}_{\rho
_{0}}^{\rho _{1}}\tilde{\omega}_{\rho _{1}\tau _{1}}D_{\lambda _{1}}^{\tau
_{1}}\tilde{\omega}^{\lambda _{1}\sigma _{1}}Z_{\sigma _{1}}^{\sigma _{0}}.
\label{210y}
\end{equation}%
Taking into account the results of Theorem \ref{th2} (see (\ref{a180})) and (%
\ref{110w})), we arrive at the relation%
\begin{equation}
\bar{A}_{\rho _{0}}^{\rho _{1}}\tilde{\omega}_{\rho _{1}\tau _{1}}D_{\lambda
_{1}}^{\tau _{1}}\tilde{\omega}^{\lambda _{1}\sigma _{1}}Z_{\sigma
_{1}}^{\sigma _{0}}\approx \bar{A}_{\rho _{0}}^{\rho _{1}}D_{\rho
_{1}}^{\sigma _{1}}Z_{\sigma _{1}}^{\sigma _{0}}\approx \bar{A}_{\rho
_{0}}^{\rho _{1}}Z_{\rho _{1}}^{\sigma _{0}},  \label{21w}
\end{equation}%
which substituted into (\ref{210y}) leads us to the formula
\begin{equation}
\mu _{\rho _{0}\lambda _{0}}^{\left( 3\right) }\mu ^{\left( 3\right) \lambda
_{0}\sigma _{0}}\approx \delta _{\rho _{0}}^{\sigma _{0}},  \label{2100q}
\end{equation}%
proving that the matrix of elements $\mu ^{\left( 3\right) \lambda
_{0}\sigma _{0}}$ given by (\ref{210z}) is indeed invertible. This proves
the theorem.$\Box $

\subsection{Irreducible approach\label{abordIRED}}

\subsubsection{Intermediate system}

Now, we introduce some new variables, $\left( y_{\alpha _{1}}\right)
_{\alpha _{1}=\overline{1,M_{1}}}$ and $\left( y_{\alpha _{3}}\right)
_{\alpha _{3}=\overline{1,M_{3}}}$, with the Poisson brackets
\begin{equation}
\left[ y_{\alpha _{1}},y_{\beta _{1}}\right] =\omega _{\alpha _{1}\beta
_{1}},\qquad \left[ y_{\alpha _{3}},y_{\beta _{3}}\right] =\omega _{\alpha
_{3}\beta _{3}},\qquad \left[ y_{\alpha _{1}},y_{\alpha _{3}}\right] =0,
\label{h25}
\end{equation}%
where $\omega _{\alpha _{1}\beta _{1}}$ and $\omega _{\alpha _{3}\beta _{3}}$
are the elements of some antisymmetric, invertible matrices, and consider a
system subject to the reducible second-class constraints
\begin{equation}
\chi _{\alpha _{0}}\approx 0,\qquad y_{\alpha _{1}}\approx 0,\qquad
y_{\alpha _{3}}\approx 0.  \label{h26}
\end{equation}%
In what follows we will call the system subject to constraints (\ref{h26})
\textquotedblleft intermediate system\textquotedblright . The Dirac bracket
on the phase-space locally described by $\left( z^{a},y_{\alpha
_{1}},y_{\alpha _{3}}\right) $ constructed with respect to the above
second-class constraints reads as
\begin{eqnarray}
\left. \left[ F,G\right] ^{\left( 3\right) \ast }\right\vert _{z,y} &=&\left[
F,G\right] -\left[ F,\chi _{\alpha _{0}}\right] \mu ^{\left( 3\right) \alpha
_{0}\beta _{0}}\left[ \chi _{\beta _{0}},G\right]  \notag \\
&&-\left[ F,y_{\alpha _{1}}\right] \omega ^{\alpha _{1}\beta _{1}}\left[
y_{\beta _{1}},G\right] -\left[ F,y_{\alpha _{3}}\right] \omega ^{\alpha
_{3}\beta _{3}}\left[ y_{\beta _{3}},G\right] ,  \label{h27}
\end{eqnarray}%
where the Poisson brackets from the right-hand side of (\ref{h27}) contain
derivatives with respect to all the variables $z^{a}$, $y_{\alpha _{1}}$,
and $y_{\alpha _{3}}$. The notations $\omega ^{\alpha _{1}\beta _{1}}$ and $%
\omega ^{\alpha _{3}\beta _{3}}$ denote the elements of the inverses of the
matrices of elements $\omega _{\alpha _{1}\beta _{1}}$ and $\omega ^{\alpha
_{3}\beta _{3}}$ respectively. The most general form of a function defined
on the phase-space of coordinates $\left( z^{a},y_{\alpha _{1}},y_{\alpha
_{3}}\right) $ is given by
\begin{equation}
F\left( z^{a},y^{A}\right) =F_{0}\left( z^{a}\right) +\int\limits_{0}^{1}%
\frac{dF(z^{a},\lambda y_{A})}{d\lambda }d\lambda =F_{0}\left( z^{a}\right)
+y_{A}G^{A}\left( z^{a},y_{B}\right) ,  \label{h27tr}
\end{equation}%
where $y_{A}=(y_{\alpha _{1}},y_{\alpha _{3}})$, $F_{0}\left( z^{a}\right)
=F\left( z^{a},0\right) $, and
\begin{equation*}
G^{A}\left( z^{a},y_{B}\right) =\int\limits_{0}^{1}\frac{\partial F\left(
z^{a},\lambda y_{B}\right) }{\partial \left( \lambda y_{A}\right) }d\lambda .
\end{equation*}%
By inserting (\ref{h27tr}) in (\ref{h27}) we obtain
\begin{equation}
\left[ F,G\right] ^{\left( 3\right) \ast }\approx \left[ F_{0},G_{0}\right]
^{\left( 3\right) \ast },  \label{h27qf}
\end{equation}%
where the previous weak equality holds on the surface (\ref{h26}). Moreover,
equations (\ref{1}) and (\ref{h26}) describe the same surface, but embedded
in two phase-spaces of different dimensions. In other words, equations (\ref%
{1}) and (\ref{h26}) represent equivalent descriptions of one and the same
constraint surface. For this reason, we will maintain the symbol of weak
equality with respect to both descriptions\footnote{%
Obviously, it is understood that we employ description (\ref{1}) whenever we
work with functions defined on the phase-space of local coordinates $z^{a}$,
but we use representation (\ref{h26}) in relation with the functions defined
on the phase-space of local coordinates $\left( z^{a},y_{\alpha
_{1}},y_{\alpha _{3}}\right) $.}. Substituting (\ref{h27tr}) in (\ref{h27})
and taking into account (\ref{h27qf}), we infer%
\begin{equation}
\left. \left[ F,G\right] ^{\left( 3\right) \ast }\right\vert _{z,y}\approx %
\left[ F,G\right] ^{\left( 3\right) \ast }.  \label{h28}
\end{equation}%
We recall that the Dirac bracket $\left[ F,G\right] ^{\left( 3\right) \ast }$
contains only derivatives with respect to the variables $z^{a}$.

\subsubsection{Irreducible system\label{sisired}}

Let $\hat{e}_{\sigma _{2}}^{\alpha _{2}}$ be the elements of an invertible
matrix, taken such that%
\begin{equation}
\bar{A}_{\alpha _{1}}^{\alpha _{2}}=A_{\alpha _{1}}^{\sigma _{2}}\hat{e}%
_{\sigma _{2}}^{\alpha _{2}},  \label{s2}
\end{equation}%
with%
\begin{equation}
A_{\alpha _{1}}^{\alpha _{2}}=\sigma _{\alpha _{1}\lambda _{1}}Z_{\beta
_{2}}^{\lambda _{1}}\sigma ^{\beta _{2}\alpha _{2}},  \label{s3}
\end{equation}%
where $\sigma _{\alpha _{1}\lambda _{1}}$ and $\sigma ^{\alpha _{2}\beta
_{2}}$ determine some invertible matrices. From (\ref{s2}) it is easy to see
that
\begin{equation}
A_{\alpha _{1}}^{\alpha _{2}}=\bar{A}_{\alpha _{1}}^{\sigma _{2}}\hat{E}%
_{\sigma _{2}}^{\alpha _{2}},  \label{s1}
\end{equation}%
with $\hat{E}_{\sigma _{2}}^{\alpha _{2}}$ the elements of the inverse of
the matrix of elements $\hat{e}_{\sigma _{2}}^{\alpha _{2}}$. Substituting (%
\ref{s2}) in (\ref{waq12}) and taking into account the invertibility of the
matrix of elements $\hat{e}_{\sigma _{2}}^{\alpha _{2}}$, we obtain%
\begin{equation}
\bar{A}_{\alpha _{0}}^{\alpha _{1}}A_{\alpha _{1}}^{\alpha _{2}}\approx 0.
\label{s4}
\end{equation}%
Next, we add an invertible matrix, whose elements will be denoted by $\hat{E}%
_{\alpha _{1}}^{\gamma _{1}}$, through the relations%
\begin{equation}
\tilde{\omega}_{\alpha _{1}\beta _{1}}=\hat{E}_{\alpha _{1}}^{\gamma
_{1}}\omega _{\gamma _{1}\lambda _{1}}\hat{E}_{\beta _{1}}^{\lambda _{1}},
\label{s5}
\end{equation}%
and define the functions
\begin{equation}
A_{\sigma _{0}}^{\rho _{1}}=\bar{A}_{\sigma _{0}}^{\alpha _{1}}\hat{E}%
_{\alpha _{1}}^{\rho _{1}}.  \label{s6}
\end{equation}%
Then, it is clear that
\begin{equation}
\tilde{\omega}^{\alpha _{1}\beta _{1}}=\hat{e}_{\sigma _{1}}^{\alpha
_{1}}\omega ^{\sigma _{1}\tau _{1}}\hat{e}_{\tau _{1}}^{\beta _{1}},
\label{s7}
\end{equation}%
with $\hat{e}_{\sigma _{1}}^{\alpha _{1}}$ the elements of the inverse of $%
\hat{E}_{\alpha _{1}}^{\gamma _{1}}$, while (\ref{s6}) produces
\begin{equation}
\bar{A}_{\sigma _{0}}^{\alpha _{1}}=A_{\sigma _{0}}^{\rho _{1}}\hat{e}_{\rho
_{1}}^{\alpha _{1}}.  \label{s8}
\end{equation}%
In this context the next theorem is shown to hold.

\begin{theorem}
\label{th4} The elements $\hat{e}_{\sigma _{1}}^{\alpha _{1}}$ and $\hat{E}%
_{\beta _{1}}^{\tau _{1}}$ can be taken such that
\begin{equation}
\hat{E}_{\sigma _{1}}^{\alpha _{1}}D_{\tau _{1}}^{\sigma _{1}}\hat{e}_{\beta
_{1}}^{\tau _{1}}\approx D_{\beta _{1}}^{\alpha _{1}}.  \label{s9}
\end{equation}
\end{theorem}

\textbf{Proof. }We take $\hat{E}_{\beta _{1}}^{\alpha _{1}}$ and $\hat{e}%
_{\beta _{1}}^{\alpha _{1}}$ such that the following relations are
satisfied:
\begin{eqnarray}
&&A_{\alpha _{0}}^{\alpha _{1}}=\sigma _{\alpha _{0}\beta _{0}}Z_{\beta
_{1}}^{\beta _{0}}\sigma ^{\beta _{1}\alpha _{1}},  \label{s10} \\
&&\sigma _{\alpha _{1}\gamma _{1}}\hat{e}_{\delta _{1}}^{\gamma _{1}}\sigma
^{\delta _{1}\beta _{1}}=\hat{e}_{\alpha _{1}}^{\beta _{1}},  \label{s101}
\end{eqnarray}%
where the matrix of elements $\sigma _{\alpha _{0}\beta _{0}}$ is taken to
be invertible and $\sigma ^{\beta _{1}\alpha _{1}}$ are the elements of the
inverse of the matrix of elements $\sigma _{\alpha _{1}\lambda _{1}}$. By
`solving' (\ref{s3}) and (\ref{s10}) with respect to the reducibility
functions of order one and two
\begin{equation}
Z_{\alpha _{1}}^{\alpha _{0}}=\sigma ^{\alpha _{0}\beta _{0}}A_{\beta
_{0}}^{\beta _{1}}\sigma _{\beta _{1}\alpha _{1}},\qquad Z_{\lambda
_{2}}^{\lambda _{1}}=\sigma ^{\lambda _{1}\tau _{1}}A_{\tau _{1}}^{\tau
_{2}}\sigma _{\tau _{2}\lambda _{2}},  \label{s11}
\end{equation}%
where $\sigma ^{\alpha _{0}\beta _{0}}$ and $\sigma _{\lambda _{2}\tau _{2}}$
are the elements of the inverses of the matrices of elements $\sigma
_{\alpha _{0}\beta _{0}}$ and $\sigma ^{\alpha _{2}\beta _{2}}$
respectively, we can write
\begin{equation}
Z_{\alpha _{1}}^{\alpha _{0}}\hat{e}_{\lambda _{1}}^{\alpha _{1}}Z_{\lambda
_{2}}^{\lambda _{1}}=\sigma ^{\alpha _{0}\beta _{0}}A_{\beta _{0}}^{\beta
_{1}}\sigma _{\beta _{1}\alpha _{1}}\hat{e}_{\lambda _{1}}^{\alpha
_{1}}\sigma ^{\lambda _{1}\tau _{1}}A_{\tau _{1}}^{\tau _{2}}\sigma _{\tau
_{2}\lambda _{2}}.  \label{s12}
\end{equation}%
From (\ref{s12}) and taking into account (\ref{s8}) and (\ref{s101}), we
deduce the relation%
\begin{equation}
Z_{\alpha _{1}}^{\alpha _{0}}\hat{e}_{\lambda _{1}}^{\alpha _{1}}Z_{\lambda
_{2}}^{\lambda _{1}}=\sigma ^{\alpha _{0}\beta _{0}}\bar{A}_{\beta
_{0}}^{\beta _{1}}A_{\beta _{1}}^{\tau _{2}}\sigma _{\tau _{2}\lambda _{2}}.
\label{s14}
\end{equation}%
Inserting now (\ref{s4}) in (\ref{s14}), we arrive at
\begin{equation}
Z_{\alpha _{1}}^{\alpha _{0}}\hat{e}_{\lambda _{1}}^{\alpha _{1}}Z_{\lambda
_{2}}^{\lambda _{1}}\approx 0.  \label{s15}
\end{equation}%
Based on the results expressed by (\ref{s4}) and (\ref{s15}), we are able
now to prove the validity of (\ref{s9}). If we make the notation
\begin{equation}
\hat{D}_{\beta _{1}}^{\alpha _{1}}=\hat{e}_{\sigma _{1}}^{\alpha
_{1}}D_{\tau _{1}}^{\sigma _{1}}\hat{E}_{\beta _{1}}^{\tau _{1}},
\label{s16}
\end{equation}%
then it is easy to see that $\hat{D}_{\beta _{1}}^{\alpha _{1}}$ is a
`projection'
\begin{equation}
\hat{D}_{\beta _{1}}^{\alpha _{1}}\hat{D}_{\lambda _{1}}^{\beta _{1}}\approx
\hat{D}_{\lambda _{1}}^{\alpha _{1}}.  \label{s17}
\end{equation}%
On the other hand, with the help of relations (\ref{s3}) and (\ref{s10}), we
deduce that $A_{\alpha _{0}}^{\alpha _{1}}A_{\alpha _{1}}^{\alpha
_{2}}\approx 0$, which further implies
\begin{equation}
A_{\alpha _{0}}^{\alpha _{1}}\bar{A}_{\alpha _{1}}^{\alpha _{2}}\approx 0,
\label{s18}
\end{equation}%
and hence we find
\begin{equation}
\bar{A}_{\alpha _{0}}^{\beta _{1}}\hat{D}_{\beta _{1}}^{\alpha _{1}}\approx
\bar{A}_{\alpha _{0}}^{\alpha _{1}}.  \label{s19}
\end{equation}%
Applying $Z_{\alpha _{1}}^{\alpha _{0}}$ on (\ref{s16}) and relying on (\ref%
{s15}), we get%
\begin{equation}
Z_{\alpha _{1}}^{\alpha _{0}}\hat{D}_{\beta _{1}}^{\alpha _{1}}\approx
Z_{\beta _{1}}^{\alpha _{0}}.  \label{s20}
\end{equation}%
Multiplying (\ref{s19}) with $Z_{\rho _{1}}^{\alpha _{0}}$ and (\ref{s20})
with $\bar{A}_{\alpha _{0}}^{\alpha _{1}}$, we are led to%
\begin{equation}
\hat{D}_{\beta _{1}}^{\alpha _{1}}D_{\rho _{1}}^{\beta _{1}}\approx D_{\rho
_{1}}^{\alpha _{1}},\qquad D_{\beta _{1}}^{\alpha _{1}}\hat{D}_{\rho
_{1}}^{\beta _{1}}\approx D_{\rho _{1}}^{\alpha _{1}}.  \label{s21}
\end{equation}%
The general solution to equations (\ref{s21}) is of the form
\begin{equation}
\hat{D}_{\beta _{1}}^{\alpha _{1}}\approx D_{\beta _{1}}^{\alpha _{1}}+\bar{A%
}_{\beta _{1}}^{\tau _{2}}M_{\tau _{2}}^{\lambda _{2}}Z_{\lambda
_{2}}^{\alpha _{1}},  \label{s22}
\end{equation}%
for an arbitrary matrix of elements $M_{\tau _{2}}^{\lambda _{2}}$. Direct
computation yields
\begin{equation}
\hat{D}_{\lambda _{1}}^{\alpha _{1}}\hat{D}_{\beta _{1}}^{\lambda
_{1}}\approx D_{\beta _{1}}^{\alpha _{1}}+\bar{A}_{\beta _{1}}^{\tau
_{2}}M_{\tau _{2}}^{\lambda _{2}}D_{\lambda _{2}}^{\rho _{2}}M_{\rho
_{2}}^{\delta _{2}}Z_{\delta _{2}}^{\alpha _{1}}.  \label{s23}
\end{equation}%
Comparing (\ref{s23}) with (\ref{s17}) and employing (\ref{s22}), we obtain
that the elements $M_{\tau _{2}}^{\lambda _{2}}$ are subject to the
equations
\begin{equation}
\bar{A}_{\beta _{1}}^{\tau _{2}}M_{\tau _{2}}^{\lambda _{2}}D_{\lambda
_{2}}^{\rho _{2}}M_{\rho _{2}}^{\delta _{2}}Z_{\delta _{2}}^{\alpha
_{1}}\approx \bar{A}_{\beta _{1}}^{\tau _{2}}M_{\tau _{2}}^{\lambda
_{2}}Z_{\lambda _{2}}^{\alpha _{1}}.  \label{s24}
\end{equation}%
It is easy to see that equations (\ref{s24}) possess two types of solutions,
namely%
\begin{equation}
M_{\tau _{2}}^{\lambda _{2}}=0,  \label{s25}
\end{equation}%
and
\begin{equation}
M_{\tau _{2}}^{\lambda _{2}}=D_{\tau _{2}}^{\lambda _{2}}.  \label{s26}
\end{equation}%
If we employ solution (\ref{s25})\footnote{%
The other solution, (\ref{s26}), produces the equation $\hat{e}_{\sigma
_{1}}^{\alpha _{1}}D_{\tau _{1}}^{\sigma _{1}}\hat{E}_{\beta _{1}}^{\tau
_{1}}\approx \delta _{\beta _{1}}^{\alpha _{1}}$, which further implies the
relation $D_{\beta _{1}}^{\sigma _{1}}\approx \delta _{\beta _{1}}^{\alpha
_{1}}$, contradicting thus (\ref{11fa}).}, from (\ref{s22}) we infer
\begin{equation}
\hat{D}_{\beta _{1}}^{\alpha _{1}}\approx D_{\beta _{1}}^{\alpha _{1}},
\label{s27}
\end{equation}%
such that (\ref{s9}) is valid. This proves the theorem. $\Box $

Replacing (\ref{s5}) and (\ref{s7}) in (\ref{a180}) and recalling (\ref{s9}%
), it is easy to obtain the relation
\begin{equation}
\omega ^{\alpha _{1}\tau _{1}}D_{\tau _{1}}^{\sigma _{1}}\omega _{\sigma
_{1}\beta _{1}}\approx D_{\beta _{1}}^{\alpha _{1}}.  \label{h27wp}
\end{equation}%
On the other hand, formulae (\ref{s5})--(\ref{s7}) imply that $\mu ^{\left(
3\right) \lambda _{0}\sigma _{0}}$ and $\mu _{\sigma _{0}\rho _{0}}^{\left(
3\right) }$ given by (\ref{210z}) and (\ref{210x}) respectively can be
expressed as
\begin{eqnarray}
\mu ^{\left( 3\right) \lambda _{0}\sigma _{0}} &\approx &M^{\left( 3\right)
\lambda _{0}\sigma _{0}}+Z_{\lambda _{1}}^{\lambda _{0}}\hat{e}_{\sigma
_{1}}^{\lambda _{1}}\omega ^{\sigma _{1}\tau _{1}}\hat{e}_{\tau
_{1}}^{\gamma _{1}}Z_{\gamma _{1}}^{\sigma _{0}},  \label{h27z} \\
\mu _{\sigma _{0}\rho _{0}}^{\left( 3\right) } &\approx &C_{\sigma _{0}\rho
_{0}}^{\left( 3\right) }+A_{\sigma _{0}}^{\rho _{1}}\omega _{\rho _{1}\tau
_{1}}A_{\rho _{0}}^{\tau _{1}}.  \label{h27x}
\end{eqnarray}%
At this point, we construct the constraints
\begin{eqnarray}
\tilde{\chi}_{\alpha _{0}} &\equiv &\chi _{\alpha _{0}}+A_{\alpha
_{0}}^{\alpha _{1}}y_{\alpha _{1}}\approx 0,  \label{h28x} \\
\tilde{\chi}_{\alpha _{2}} &\equiv &Z_{\alpha _{2}}^{\alpha _{1}}y_{\alpha
_{1}}+A_{\alpha _{2}}^{\alpha _{3}}y_{\alpha _{3}}\approx 0.  \label{h28x2}
\end{eqnarray}%
Under these considerations, we are able to prove the following key theorem.

\begin{theorem}
\label{th5} Constraints (\ref{h28x}) and (\ref{h28x2}) satisfy the following
properties:

\noindent (i) equivalence to (\ref{h26}), i.e.\footnote{%
Due to the equivalence expressed by (\ref{h28y}), in the following we will
use the same symbol of weak equality in relation to both the constraints (%
\ref{h26}) and (\ref{h28x})--(\ref{h28x2}) respectively.}
\begin{equation}
\left( \tilde{\chi}_{\alpha _{0}}\approx 0,\tilde{\chi}_{\alpha _{2}}\approx
0\right) \Leftrightarrow \left( \chi _{\alpha _{0}}\approx 0,y_{\alpha
_{1}}\approx 0,y_{\alpha _{3}}\approx 0\right) ;  \label{h28y}
\end{equation}%
(ii) second-class behaviour, i.e. the matrix of elements%
\begin{equation}
C_{\Delta \Delta ^{\prime }}=\left[ \tilde{\chi}_{\Delta },\tilde{\chi}%
_{\Delta ^{\prime }}\right]  \label{h28z}
\end{equation}%
is invertible, where%
\begin{equation}
\tilde{\chi}_{\Delta }\equiv \left( \tilde{\chi}_{\alpha _{0}},\tilde{\chi}%
_{\alpha _{2}}\right) ;  \label{bv1}
\end{equation}%
(iii) irreducibility.
\end{theorem}

\textbf{Proof. }(i) It is easy to see that if (\ref{h26}) hold, then (\ref%
{h28x}) and (\ref{h28x2}) also hold
\begin{equation}
\left( \chi _{\alpha _{0}}\approx 0,y_{\alpha _{1}}\approx 0,y_{\alpha
_{3}}\approx 0\right) \Rightarrow \left( \tilde{\chi}_{\alpha _{0}}\approx 0,%
\tilde{\chi}_{\alpha _{2}}\approx 0\right) .  \label{wq1}
\end{equation}%
On the other hand, from (\ref{h28x}) and (\ref{h28x2}) one can express $\chi
_{\alpha _{0}}$, $y_{\alpha _{1}}$, and $y_{\alpha _{3}}$ in terms of $%
\tilde{\chi}_{\alpha _{0}}$ and $\tilde{\chi}_{\alpha _{2}}$ of the form%
\begin{equation}
\chi _{\alpha _{0}}=D_{\alpha _{0}}^{\beta _{0}}\tilde{\chi}_{\beta
_{0}},\qquad y_{\alpha _{1}}=\hat{e}_{\alpha _{1}}^{\gamma _{1}}Z_{\gamma
_{1}}^{\alpha _{0}}\tilde{\chi}_{\alpha _{0}}+\bar{A}_{\alpha _{1}}^{\alpha
_{2}}\tilde{\chi}_{\alpha _{2}},\qquad y_{\alpha _{3}}=\bar{D}_{\alpha
_{3}}^{\gamma _{3}}Z_{\gamma _{3}}^{\alpha _{2}}\tilde{\chi}_{\alpha _{2}}.
\label{h36}
\end{equation}%
Using (\ref{h36}), it follows that if (\ref{h28x}) and (\ref{h28x2}) hold,
then (\ref{h26}) hold, too
\begin{equation}
\left( \tilde{\chi}_{\alpha _{0}}\approx 0,\tilde{\chi}_{\alpha _{2}}\approx
0\right) \Rightarrow \left( \chi _{\alpha _{0}}\approx 0,y_{\alpha
_{1}}\approx 0,y_{\alpha _{3}}\approx 0\right) .  \label{h37}
\end{equation}%
Relations (\ref{wq1}) and (\ref{h37}) prove (i).

(ii) With the help of formulae (\ref{h28x}) and (\ref{h28x2}), we find the
expressions of the Poisson brackets among the functions $\tilde{\chi}%
_{\Delta }$ as:%
\begin{eqnarray}
&&\left[ \tilde{\chi}_{\alpha _{0}},\tilde{\chi}_{\beta _{0}}\right] \approx
\mu _{\alpha _{0}\beta _{0}}^{\left( 3\right) },\qquad \left[ \tilde{\chi}%
_{\alpha _{0}},\tilde{\chi}_{\beta _{2}}\right] \approx A_{\alpha
_{0}}^{\alpha _{1}}\omega _{\alpha _{1}\beta _{1}}Z_{\beta _{2}}^{\beta
_{1}},  \label{c62} \\
&&\left[ \tilde{\chi}_{\alpha _{2}},\tilde{\chi}_{\beta _{2}}\right] \approx
Z_{\alpha _{2}}^{\alpha _{1}}\omega _{\alpha _{1}\beta _{1}}Z_{\beta
_{2}}^{\beta _{1}}+A_{\alpha _{2}}^{\alpha _{3}}\omega _{\alpha _{3}\beta
_{3}}A_{\beta _{2}}^{\beta _{3}},  \label{c64}
\end{eqnarray}%
where $\mu _{\alpha _{0}\beta _{0}}^{\left( 3\right) }$ reads as in (\ref%
{h27x}). Then, the matrix of their Poisson brackets, of elements $C_{\Delta
\Delta ^{\prime }}$, takes the concrete form
\begin{equation}
C_{\Delta \Delta ^{\prime }}=\left(
\begin{array}{cc}
\mu _{\alpha _{0}\beta _{0}}^{\left( 3\right) } & A_{\alpha _{0}}^{\alpha
_{1}}\omega _{\alpha _{1}\beta _{1}}Z_{\beta _{2}}^{\beta _{1}} \\
Z_{\alpha _{2}}^{\alpha _{1}}\omega _{\alpha _{1}\beta _{1}}A_{\beta
_{0}}^{\beta _{1}} & \phi _{\alpha _{2}\beta _{2}}^{(3)}%
\end{array}%
\right) ,  \label{h29z}
\end{equation}%
where $\Delta =\left( \alpha _{0},\alpha _{2}\right) $ indexes the line, $%
\Delta ^{\prime }=\left( \beta _{0},\beta _{2}\right) $ the column, and $%
\phi _{\alpha _{2}\beta _{2}}^{(3)}$ means%
\begin{equation}
\phi _{\alpha _{2}\beta _{2}}^{(3)}=Z_{\alpha _{2}}^{\alpha _{1}}\omega
_{\alpha _{1}\beta _{1}}Z_{\beta _{2}}^{\beta _{1}}+A_{\alpha _{2}}^{\alpha
_{3}}\omega _{\alpha _{3}\beta _{3}}A_{\beta _{2}}^{\beta _{3}}.
\label{h29z1}
\end{equation}%
In order to prove the invertibility of the matrix (\ref{h29z}), we will give
its inverse. Direct computation shows that the matrix%
\begin{equation}
C^{\Delta ^{\prime }\Delta ^{\prime \prime }}=\left(
\begin{array}{cc}
\mu ^{\left( 3\right) \beta _{0}\rho _{0}} & Z_{\gamma _{1}}^{\beta _{0}}%
\hat{e}_{\sigma _{1}}^{\gamma _{1}}\omega ^{\sigma _{1}\lambda _{1}}\bar{A}%
_{\lambda _{1}}^{\rho _{2}} \\
\bar{A}_{\sigma _{1}}^{\beta _{2}}\omega ^{\sigma _{1}\lambda _{1}}\hat{e}%
_{\lambda _{1}}^{\gamma _{1}}Z_{\gamma _{1}}^{\rho _{0}} & \psi ^{\left(
3\right) \beta _{2}\rho _{2}}%
\end{array}%
\right) ,  \label{h29y}
\end{equation}%
with $\mu ^{\left( 3\right) \beta _{0}\rho _{0}}$ given by (\ref{h27x}) and $%
\psi ^{\left( 3\right) \beta _{2}\rho _{2}}$ of the form%
\begin{equation}
\psi ^{\left( 3\right) \beta _{2}\rho _{2}}=\bar{A}_{\sigma _{1}}^{\beta
_{2}}\omega ^{\sigma _{1}\lambda _{1}}\bar{A}_{\lambda _{1}}^{\rho
_{2}}+Z_{\tau _{3}}^{\beta _{2}}\bar{D}_{\gamma _{3}}^{\tau _{3}}\omega
^{\gamma _{3}\lambda _{3}}\bar{D}_{\lambda _{3}}^{\sigma _{3}}Z_{\sigma
_{3}}^{\rho _{2}},  \label{h31}
\end{equation}%
satisfies the relations
\begin{equation}
C_{\Delta \Delta ^{\prime }}C^{\Delta ^{\prime }\Delta ^{\prime \prime
}}\approx \left(
\begin{array}{cc}
\delta _{\alpha _{0}}^{\rho _{0}} & \mathbf{0} \\
\mathbf{0} & \delta _{\alpha _{2}}^{\rho _{2}}%
\end{array}%
\right) ,  \label{h30}
\end{equation}%
so it is indeed the inverse of (\ref{h29y}). This proves (ii).

(iii) Since matrix (\ref{h29z}) is invertible, it follows that it possesses
no nontrivial null vectors and hence the functions $\tilde{\chi}_{\Delta }$
are independent, which is equivalent to the fact that the constraint set
given by (\ref{h28x}) and (\ref{h28x2}) is irreducible. This proves (iii). $%
\Box $

Taking into account the result (\ref{h29y}), the Dirac bracket built with
respect to the irreducible second-class constraint set (\ref{h28x}) and (\ref%
{h28x2})
\begin{equation}
\left. \left[ F,G\right] ^{\left( 3\right) \ast }\right\vert _{\mathrm{ired}%
}=\left[ F,G\right] -\left[ F,\tilde{\chi}_{\Delta }\right] C^{\Delta \Delta
^{\prime }}\left[ \tilde{\chi}_{\Delta ^{\prime }},G\right] ,  \label{i40}
\end{equation}%
takes the concrete form
\begin{eqnarray}
\left. \left[ F,G\right] ^{\left( 3\right) \ast }\right\vert _{\mathrm{ired}%
} &=&\left[ F,G\right] -\left[ F,\tilde{\chi}_{\alpha _{0}}\right] \mu
^{\left( 3\right) \alpha _{0}\beta _{0}}\left[ \tilde{\chi}_{\beta _{0}},G%
\right]  \notag \\
&&-\left[ F,\tilde{\chi}_{\alpha _{0}}\right] Z_{\gamma _{1}}^{\alpha _{0}}%
\hat{e}_{\sigma _{1}}^{\gamma _{1}}\omega ^{\sigma _{1}\lambda _{1}}\bar{A}%
_{\lambda _{1}}^{\beta _{2}}\left[ \tilde{\chi}_{\beta _{2}},G\right]  \notag
\\
&&-\left[ F,\tilde{\chi}_{\alpha _{2}}\right] \bar{A}_{\sigma _{1}}^{\alpha
_{2}}\omega ^{\sigma _{1}\lambda _{1}}\hat{e}_{\lambda _{1}}^{\gamma
_{1}}Z_{\gamma _{1}}^{\beta _{0}}\left[ \tilde{\chi}_{\beta _{0}},G\right]
\notag \\
&&-\left[ F,\tilde{\chi}_{\alpha _{2}}\right] \left( \bar{A}_{\sigma
_{1}}^{\alpha _{2}}\omega ^{\sigma _{1}\lambda _{1}}\bar{A}_{\lambda
_{1}}^{\beta _{2}}\right.  \notag \\
&&\left. +Z_{\tau _{3}}^{\alpha _{2}}\bar{D}_{\gamma _{3}}^{\tau _{3}}\omega
^{\gamma _{3}\lambda _{3}}\bar{D}_{\lambda _{3}}^{\sigma _{3}}Z_{\sigma
_{3}}^{\beta _{2}}\right) \left[ \tilde{\chi}_{\beta _{2}},G\right] .
\label{c65}
\end{eqnarray}

\begin{theorem}
\label{th6} The Dirac bracket with respect to the irreducible second-class
constraints (\ref{c65}) coincides with that of the intermediate system
\begin{equation}
\left. \left[ F,G\right] ^{\left( 3\right) \ast }\right\vert _{\mathrm{ired}%
}\approx \left. \left[ F,G\right] ^{\left( 3\right) \ast }\right\vert _{z,y}.
\label{h32y}
\end{equation}
\end{theorem}

\textbf{Proof. }In order to prove this theorem, we start from the right-hand
side of (\ref{c65}) and show that it is (weakly) equal with the right-hand
side of (\ref{h27}). Collecting the results expressed by relations (\ref{2}%
), (\ref{d2}), (\ref{10qa}), (\ref{11fa}), (\ref{1qaa}), (\ref{a70}), (\ref%
{s6}), (\ref{s9}), (\ref{h27z}), (\ref{h28x}), and (\ref{h28x2}), by direct
computation we obtain:
\begin{eqnarray}
&&\left[ F,\tilde{\chi}_{\alpha _{0}}\right] \mu ^{\left( 3\right) \alpha
_{0}\beta _{0}}\left[ \tilde{\chi}_{\beta _{0}},G\right] \approx  \notag \\
&&\left[ F,\chi _{\alpha _{0}}\right] \mu ^{\left( 3\right) \alpha _{0}\beta
_{0}}\left[ \chi _{\beta _{0}},G\right] +\left[ F,y_{\alpha _{1}}\right]
D_{\sigma _{1}}^{\alpha _{1}}\omega ^{\sigma _{1}\lambda _{1}}D_{\lambda
_{1}}^{\beta _{1}}\left[ y_{\beta _{1}},G\right] , \\
&&\left[ F,\tilde{\chi}_{\alpha _{0}}\right] Z_{\gamma _{1}}^{\alpha _{0}}%
\hat{e}_{\sigma _{1}}^{\gamma _{1}}\omega ^{\sigma _{1}\lambda _{1}}\bar{A}%
_{\lambda _{1}}^{\beta _{2}}\left[ \tilde{\chi}_{\beta _{2}},G\right] \approx
\notag \\
&&\left[ F,y_{\alpha _{1}}\right] D_{\sigma _{1}}^{\alpha _{1}}\omega
^{\sigma _{1}\lambda _{1}}\left( \delta _{\lambda _{1}}^{\beta
_{1}}-D_{\lambda _{1}}^{\beta _{1}}\right) \left[ y_{\beta _{1}},G\right] ,
\\
&&\left[ F,\tilde{\chi}_{\alpha _{2}}\right] \bar{A}_{\sigma _{1}}^{\alpha
_{2}}\omega ^{\sigma _{1}\lambda _{1}}\hat{e}_{\lambda _{1}}^{\gamma
_{1}}Z_{\gamma _{1}}^{\beta _{0}}\left[ \tilde{\chi}_{\beta _{0}},G\right]
\approx  \notag \\
&&\left[ F,y_{\alpha _{1}}\right] \left( \delta _{\sigma _{1}}^{\alpha
_{1}}-D_{\sigma _{1}}^{\alpha _{1}}\right) \omega ^{\sigma _{1}\lambda
_{1}}D_{\lambda _{1}}^{\beta _{1}}\left[ y_{\beta _{1}},G\right] , \\
&&\left[ F,\tilde{\chi}_{\alpha _{2}}\right] \left( \bar{A}_{\sigma
_{1}}^{\alpha _{2}}\omega ^{\sigma _{1}\lambda _{1}}\bar{A}_{\lambda
_{1}}^{\beta _{2}}+Z_{\tau _{3}}^{\alpha _{2}}\bar{D}_{\gamma _{3}}^{\tau
_{3}}\omega ^{\gamma _{3}\lambda _{3}}\bar{D}_{\lambda _{3}}^{\sigma
_{3}}Z_{\sigma _{3}}^{\beta _{2}}\right) \left[ \tilde{\chi}_{\beta _{2}},G%
\right] \approx  \notag \\
&&\left[ F,y_{\alpha _{1}}\right] \left( \delta _{\sigma _{1}}^{\alpha
_{1}}-D_{\sigma _{1}}^{\alpha _{1}}\right) \omega ^{\sigma _{1}\lambda
_{1}}\left( \delta _{\lambda _{1}}^{\beta _{1}}-D_{\lambda _{1}}^{\beta
_{1}}\right) \left[ y_{\beta _{1}},G\right]  \notag \\
&&+\left[ F,y_{\alpha _{3}}\right] \omega ^{\alpha _{3}\beta _{3}}\left[
y_{\beta _{3}},G\right] ,
\end{eqnarray}%
Substituting the previous results in (\ref{c65}), we arrive precisely at (%
\ref{h32y}), which proves the theorem. $\Box $

\subsection{Basic result for $L=3$}

Combining (\ref{h28}) and (\ref{h32y}), we are led to the result%
\begin{equation}
\left[ F,G\right] ^{\left( 3\right) \ast }\approx \left. \left[ F,G\right]
^{\left( 3\right) \ast }\right\vert _{\mathrm{ired}}.  \label{h32}
\end{equation}%
The last formula proves that we can indeed approach third-order reducible
second-class constraints in an irreducible fashion.

\section{Generalization to an arbitrary reducibility order $L$\label{Lth}}

\subsection{Reducible approach\label{redL}}

In the sequel we generalize the previous results to the case of a system of
second-class constraints, reducible of an arbitrary order $L$%
\begin{equation}
Z_{\alpha _{1}}^{\alpha _{0}}\chi _{\alpha _{0}}=0,\qquad Z_{\alpha
_{2}}^{\alpha _{1}}Z_{\alpha _{1}}^{\alpha _{0}}\approx 0,\ldots ,\qquad
Z_{\alpha _{L}}^{\alpha _{L-1}}Z_{\alpha _{L-1}}^{\alpha _{L-2}}\approx 0,
\label{l1}
\end{equation}%
with $\alpha _{k}=\overline{1,M_{k}}$ for each $k=\overline{1,L}$.\ In
addition, the reducibility functions of maximum order ($L$), $Z_{\alpha
_{L}}^{\alpha _{L-1}}$, are assumed to be all independent. Consequently, the
number of independent second-class constraints is equal to $M\equiv
\sum\limits_{k=0}^{L}\left( -\right) ^{k}M_{k}$. Therefore, we can work
again here with a Dirac bracket of the type (\ref{110b}), but in terms of $M$
independent functions $\chi _{A}$, i.e.
\begin{equation}
\left[ F,G\right] ^{\left( L\right) \ast }=\left[ F,G\right] -\left[ F,\chi
_{A}\right] M^{\left( L\right) AB}\left[ \chi _{B},G\right] ,\qquad A=%
\overline{1,M},  \label{l2}
\end{equation}%
where $C_{AB}^{\left( L\right) }M^{\left( L\right) BC}\approx \delta
_{A}^{C} $, with $C_{AB}^{\left( L\right) }=\left[ \chi _{A},\chi _{B}\right]
$. The matrix of the Poisson brackets among the constraint functions%
\begin{equation}
C_{\alpha _{0}\beta _{0}}^{\left( L\right) }=\left[ \chi _{\alpha _{0}},\chi
_{\beta _{0}}\right]  \label{l3}
\end{equation}%
is not invertible due to the relations
\begin{equation}
Z_{\alpha _{1}}^{\alpha _{0}}C_{\alpha _{0}\beta _{0}}^{\left( L\right)
}\approx 0,  \label{l4}
\end{equation}%
but its rank is equal to $M$.

Just like in the case of order three of reducibility, we introduce some
functions $\left( \bar{A}_{\alpha _{k-1}}^{\alpha _{k}}\right) _{k=\overline{%
1,L}}$, subject to the relations
\begin{eqnarray*}
&&\mathrm{rank}\left( Z_{\alpha _{k}}^{\beta _{k-1}}\bar{A}_{\beta
_{k-1}}^{\gamma _{k}}\right) \approx \sum\limits_{i=k}^{L}\left( -\right)
^{k+i}M_{i}, \\
&&\bar{A}_{\alpha _{k-2}}^{\alpha _{k-1}}\bar{A}_{\alpha _{k-1}}^{\alpha
_{k}}\approx 0.
\end{eqnarray*}%
The Dirac bracket from (\ref{l2}) can be written, like in the previous
situation, in terms of all the second-class constraint functions. Going
along a line similar to that from subsection \ref{parDirac1}, we introduce
an antisymmetric matrix, of elements $M^{\left( L\right) \alpha _{0}\beta
_{0}}$, through the relation
\begin{equation}
C_{\alpha _{0}\beta _{0}}^{\left( L\right) }M^{\left( L\right) \beta
_{0}\gamma _{0}}\approx D_{\alpha _{0}}^{\gamma _{0}},  \label{l5}
\end{equation}%
such that%
\begin{equation}
\left[ F,G\right] ^{\left( L\right) \ast }=\left[ F,G\right] -\left[ F,\chi
_{\alpha _{0}}\right] M^{\left( L\right) \alpha _{0}\beta _{0}}\left[ \chi
_{\beta _{0}},G\right]  \label{l6}
\end{equation}%
defines the same Dirac bracket like (\ref{l2}) on the surface (\ref{1}).
Similar to the case of third-order reducible second-class constraints, the
Dirac bracket for $L$-order reducible constraints can be expressed in terms
of a noninvertible matrix.

\begin{theorem}
\label{th7} There exists an invertible, antisymmetric matrix $\mu ^{\left(
L\right) \alpha _{0}\beta _{0}}$ such that Dirac bracket (\ref{l6}) takes
the form
\begin{equation}
\left[ F,G\right] ^{\left( L\right) \ast }=\left[ F,G\right] -\left[ F,\chi
_{\alpha _{0}}\right] \mu ^{\left( L\right) \alpha _{0}\beta _{0}}\left[
\chi _{\beta _{0}},G\right]  \label{l7}
\end{equation}%
on the surface (\ref{1}).
\end{theorem}

The relationship between the invertible matrix $\mu ^{\left( L\right) }$ and
the matrix $M^{\left( L\right) }$ is given by a relation similar to that
from the third-order reducible case%
\begin{equation}
M^{\left( L\right) \alpha _{0}\beta _{0}}\approx D_{\lambda _{0}}^{\alpha
_{0}}\mu ^{\left( L\right) \lambda _{0}\sigma _{0}}D_{\sigma _{0}}^{\beta
_{0}}.  \label{l71}
\end{equation}

\subsection{Irreducible approach\label{iredL}}

\subsubsection{Intermediate system}

Now, we introduce some new variables, $\left( y_{\alpha _{2k+1}}\right)
_{\alpha _{2k+1}=\overline{1,M_{2k+1}}},$ with $k=\overline{0,\left[ \frac{%
L-1}{2}\right] }$, exhibiting the Poisson brackets
\begin{equation}
\left[ y_{\alpha _{i}},y_{\beta _{j}}\right] =\omega _{\alpha _{i}\beta
_{j}}\delta _{ij},  \label{l8}
\end{equation}%
where $\omega _{\alpha _{i}\beta _{j}}$ are the elements of an
antisymmetric, invertible matrix, and consider the system subject to the
reducible second-class constraints
\begin{equation}
\chi _{\alpha _{0}}\approx 0,\qquad \left( y_{\alpha _{2k+1}}\right) _{k=%
\overline{0,\left[ \frac{L-1}{2}\right] }}\approx 0.  \label{l9}
\end{equation}%
The system constrained to satisfy (\ref{l9}) will be called
\textquotedblleft intermediate system\textquotedblright\ in what follows.
The Dirac bracket on the phase-space locally parameterized by the variables $%
\left( z^{a},\left( y_{\alpha _{2k+1}}\right) _{k=\overline{0,\left[ \frac{%
L-1}{2}\right] }}\right) $, constructed with respect to the above
second-class constraints, reads as
\begin{eqnarray}
\left. \left[ F,G\right] ^{\left( L\right) \ast }\right\vert _{z,y} &=&\left[
F,G\right] -\left[ F,\chi _{\alpha _{0}}\right] \mu ^{\left( L\right) \alpha
_{0}\beta _{0}}\left[ \chi _{\beta _{0}},G\right]  \notag \\
&&-\sum\limits_{k=0}^{\left[ \frac{L-1}{2}\right] }\left[ F,y_{\alpha
_{2k+1}}\right] \omega ^{\alpha _{2k+1}\beta _{2k+1}}\left[ y_{\beta
_{2k+1}},G\right] ,  \label{l10}
\end{eqnarray}%
where the Poisson brackets from the right-hand side of (\ref{l10}) contain
derivatives with respect to all the variables $z^{a}$ and $\left( y_{\alpha
_{2k+1}}\right) _{k=\overline{0,\left[ \frac{L-1}{2}\right] }}$ and $\omega
^{\alpha _{2k+1}\beta _{2k+1}}$ denote the elements of the inverse of the
matrix of elements $\omega _{\alpha _{2k+1}\beta _{2k+1}}$. In this case the
most general form of a function defined on the phase-space locally
parameterized by $\left( z^{a},\left( y_{\alpha _{2k+1}}\right) _{k=%
\overline{0,\left[ \frac{L-1}{2}\right] }}\right) $ is given by
\begin{equation}
F\left( z^{a},y_{A}\right) =F_{0}\left( z^{a}\right) +\int\limits_{0}^{1}%
\frac{dF\left( z_{a},\lambda y_{A}\right) }{d\lambda }d\lambda =F_{0}\left(
z^{a}\right) +y_{A}G^{A}\left( z^{a},y_{B}\right) ,  \label{l11}
\end{equation}%
with $y_{A}=\left( y_{\alpha _{2k+1}}\right) _{k=\overline{0,\left[ \frac{L-1%
}{2}\right] }}$, $F_{0}\left( z^{a}\right) =F_{0}\left( z^{a},0\right) $,
and
\begin{equation*}
G^{A}\left( z^{a},y_{B}\right) =\int\limits_{0}^{1}\frac{\partial F\left(
z_{a},\lambda y_{A}\right) }{\partial \left( \lambda y_{A}\right) }d\lambda .
\end{equation*}%
If we introduce (\ref{l11}) in (\ref{l10}), then we obtain%
\begin{equation}
\left[ F,G\right] ^{\left( L\right) \ast }\approx \left[ F_{0},G_{0}\right]
^{\left( L\right) \ast },  \label{l12}
\end{equation}%
where the previous weak equality takes place on the surface defined by (\ref%
{l9}). Moreover, equations (\ref{1}) and (\ref{l9}) describe the same
surface, but embedded in phase-spaces of different dimensions, such that (%
\ref{1}) and (\ref{l9}) are equivalent descriptions of one and the same
constraint surface. This is why we will maintain the same sign of weak
equality related to both descriptions\footnote{%
It is understood that for the functions defined on the phase-space locally
parameterized by the variables $z^{a}$ we use (\ref{1}) and for those
defined on the larger phase-space, of coordinates $\left( z^{a},\left(
y_{\alpha _{2k+1}}\right) _{k=\overline{0,\left[ \frac{L-1}{2}\right] }%
}\right) $, we employ representation (\ref{l9}).}. Replacing (\ref{l11}) in (%
\ref{l10}) and making use of (\ref{l12}), we infer the result
\begin{equation}
\left. \left[ F,G\right] ^{\left( L\right) \ast }\right\vert _{z,y}\approx %
\left[ F,G\right] ^{\left( L\right) \ast }.  \label{l13}
\end{equation}%
We recall the fact that the Dirac bracket $\left[ F,G\right] ^{\left(
L\right) \ast }$ contains only derivatives with respect to the original
phase-space variables $z^{a}$.

\subsubsection{Irreducible system\label{constrIIired}}

In order to construct the irreducible system in the general case, we act in
a manner similar to that exposed in subsection \ref{sisired} and start by
adding the constraints:

\noindent -if $L$ odd%
\begin{eqnarray}
\tilde{\chi}_{\alpha _{0}} &\equiv &\chi _{\alpha _{0}}+A_{\alpha
_{0}}^{\alpha _{1}}y_{\alpha _{1}}\approx 0,  \label{l14} \\
\tilde{\chi}_{\alpha _{2k}} &\equiv &Z_{\alpha _{2k}}^{\alpha
_{2k-1}}y_{\alpha _{2k-1}}+A_{\alpha _{2k}}^{\alpha _{2k+1}}y_{\alpha
_{2k+1}}\approx 0,\qquad k=\overline{1,\left[ \frac{L}{2}\right] };
\label{l15}
\end{eqnarray}%
-if $L$ even%
\begin{eqnarray}
\tilde{\chi}_{\alpha _{0}} &\equiv &\chi _{\alpha _{0}}+A_{\alpha
_{0}}^{\alpha _{1}}y_{\alpha _{1}}\approx 0,  \label{l16} \\
\tilde{\chi}_{\alpha _{2k}} &\equiv &Z_{\alpha _{2k}}^{\alpha
_{2k-1}}y_{\alpha _{2k-1}}+A_{\alpha _{2k}}^{\alpha _{2k+1}}y_{\alpha
_{2k+1}}\approx 0,\qquad k=\overline{1,\frac{L}{2}-1},  \label{l17} \\
\tilde{\chi}_{\alpha _{L}} &\equiv &Z_{\alpha _{L}}^{\alpha _{L-1}}y_{\alpha
_{L-1}}\approx 0.  \label{l18}
\end{eqnarray}%
These constraints are defined on the larger phase-space, locally
parameterized by $\left( z^{a},\left( y_{\alpha _{2k+1}}\right) _{k=%
\overline{0,\left[ \frac{L-1}{2}\right] }}\right) $. The functions $%
A_{\alpha _{2k}}^{\alpha _{2k+1}}$ appearing in the above are defined by the
relations:

\noindent -if $L$ odd%
\begin{eqnarray}
\bar{A}_{\alpha _{2k}}^{\alpha _{2k+1}} &=&A_{\alpha _{2k}}^{\beta _{2k+1}}%
\hat{e}_{\beta _{2k+1}}^{\alpha _{2k+1}},\qquad k=\overline{0,\left[ \frac{L%
}{2}\right] -1},  \label{l19} \\
\bar{A}_{\alpha _{L-1}}^{\alpha _{L}} &=&A_{\alpha _{L-1}}^{\beta _{L}}\bar{D%
}_{\beta _{L}}^{\alpha _{L}};  \label{l20}
\end{eqnarray}%
-if $L$ even%
\begin{equation}
\bar{A}_{\alpha _{2k}}^{\alpha _{2k+1}}=A_{\alpha _{2k}}^{\beta _{2k+1}}\hat{%
e}_{\beta _{2k+1}}^{\alpha _{2k+1}},\qquad k=\overline{0,\frac{L}{2}-1}.
\label{l21}
\end{equation}%
The elements $\hat{e}_{\beta _{2k+1}}^{\alpha _{2k+1}}$ determine an
invertible matrix and $\bar{D}_{\beta _{L}}^{\alpha _{L}}$ are the elements
of the inverse of the matrix of elements $D_{\alpha _{L}}^{\beta
_{L}}=Z_{\alpha _{L}}^{\gamma _{L-1}}A_{\gamma _{L-1}}^{\beta _{L}}$.

In the following we show that (\ref{l14}) and (\ref{l15}) (or (\ref{l16})--(%
\ref{l18})) display all the desired properties: equivalence with the
intermediate system (\ref{l9}), second-class behaviour, irreducibility and,
most important, the fact that associated Dirac bracket (weakly) coincides
with the original one, corresponding to the second-order reducible
second-class constraints. The proof of all these properties is contained
within the next two theorems.

\begin{theorem}
\label{th8} Constraints (\ref{l14}) and (\ref{l15}) (or (\ref{l16})--(\ref%
{l18})) fulfill the following properties:

\noindent (i) equivalence to (\ref{l9}), i.e.
\begin{equation}
\left( \tilde{\chi}_{\alpha _{2k}}\right) _{k=\overline{0,\left[ \frac{L}{2}%
\right] }}\approx 0\Leftrightarrow \left( \chi _{\alpha _{0}}\approx
0,\left( y_{\alpha _{2k+1}}\right) _{k=\overline{0,\left[ \frac{L-1}{2}%
\right] }}\approx 0\right) ;  \label{l22}
\end{equation}%
(ii) second-class behaviour, i.e. the matrix of elements
\begin{equation}
C_{\Delta \Delta ^{\prime }}=\left[ \tilde{\chi}_{\Delta },\tilde{\chi}%
_{\Delta ^{\prime }}\right] ,  \label{l24}
\end{equation}%
is invertible, where
\begin{equation}
\tilde{\chi}_{\Delta }\equiv \left( \tilde{\chi}_{\alpha _{2k}}\right) _{k=%
\overline{0,\left[ \frac{L}{2}\right] }};  \label{l25}
\end{equation}%
(iii) irreducibility.
\end{theorem}

\textbf{Proof. }(i) It is easy to see that if (\ref{l9}) hold, then (\ref%
{l14}) and (\ref{l15}) (or (\ref{l16})--(\ref{l18})) also hold
\begin{equation}
\left( \chi _{\alpha _{0}}\approx 0,\left( y_{\alpha _{2k+1}}\right) _{k=%
\overline{0,\left[ \frac{L-1}{2}\right] }}\approx 0\right) \Rightarrow
\left( \tilde{\chi}_{\alpha _{2k}}\right) _{k=\overline{0,\left[ \frac{L}{2}%
\right] }}\approx 0.  \label{l26}
\end{equation}%
From (\ref{l14}) and (\ref{l15}) (or (\ref{l16})--(\ref{l18})) it is simple
to express the original constraint functions $\chi _{\alpha _{0}}$ and the
newly added phase-space variables $\left( y_{\alpha _{2k+1}}\right) _{k=%
\overline{0,\left[ \frac{L-1}{2}\right] }}$ in terms of $\tilde{\chi}%
_{\alpha _{0}}$ and $\left( \tilde{\chi}_{\alpha _{2k}}\right) _{k=\overline{%
1,\left[ \frac{L}{2}\right] }}$ as follows:

\noindent -if $L$ odd
\begin{eqnarray}
\chi _{\alpha _{0}} &=&D_{\alpha _{0}}^{\beta _{0}}\tilde{\chi}_{\beta _{0}},
\label{l28} \\
y_{\alpha _{2k+1}} &=&\hat{e}_{\alpha _{2k+1}}^{\beta _{2k+1}}Z_{\beta
_{2k+1}}^{\beta _{2k}}\tilde{\chi}_{\beta _{2k}}+\bar{A}_{\alpha
_{2k+1}}^{\alpha _{2k+2}}\tilde{\chi}_{\alpha _{2k+2}},\qquad k=\overline{0,%
\left[ \frac{L}{2}\right] -1},  \label{l29} \\
y_{\alpha _{L}} &=&\bar{D}_{\alpha _{L}}^{\beta _{L}}Z_{\beta _{L}}^{\beta
_{L-1}}\tilde{\chi}_{\beta _{L-1}};  \label{l30}
\end{eqnarray}%
-if $L$ even
\begin{eqnarray}
\chi _{\alpha _{0}} &=&D_{\alpha _{0}}^{\beta _{0}}\tilde{\chi}_{\beta _{0}},
\label{l31} \\
y_{\alpha _{2k+1}} &=&\hat{e}_{\alpha _{2k+1}}^{\beta _{2k+1}}Z_{\beta
_{2k+1}}^{\beta _{2k}}\tilde{\chi}_{\beta _{2k}}+\bar{A}_{\alpha
_{2k+1}}^{\alpha _{2k+2}}\tilde{\chi}_{\alpha _{2k+2}},\qquad k=\overline{0,%
\frac{L}{2}-1}.  \label{l32}
\end{eqnarray}

From (\ref{l28})--(\ref{l30}) (or (\ref{l31}) and (\ref{l32})) we obtain
that if (\ref{l14}) and (\ref{l15}) (or (\ref{l16})--(\ref{l18})) hold, then
(\ref{l9}) holds, too
\begin{equation}
\left( \tilde{\chi}_{\alpha _{0}}\approx 0,\left( \tilde{\chi}_{\alpha
_{2k}}\right) _{k=\overline{1,\left[ \frac{L}{2}\right] }}\approx 0\right)
\Rightarrow \left( \chi _{\alpha _{0}}\approx 0,\left( y_{\alpha
_{2k+1}}\right) _{k=\overline{0,\left[ \frac{L-1}{2}\right] }}\approx
0\right) .  \label{l33}
\end{equation}%
Relations (\ref{l26}) and (\ref{l33}) prove (i).

(ii) Now, we employ formulae (\ref{l14}) and (\ref{l15}) (or (\ref{l16})--(%
\ref{l18})) and find the concrete form of the Poisson brackets among the
constraint functions $\tilde{\chi}_{\Delta }$ as:

\noindent -if $L$ odd%
\begin{eqnarray}
&&\left[ \tilde{\chi}_{\alpha _{0}},\tilde{\chi}_{\beta _{0}}\right] \approx
\mu _{\alpha _{0}\beta _{0}}^{\left( L\right) },  \label{l35} \\
&&\left[ \tilde{\chi}_{\alpha _{2k}},\tilde{\chi}_{\beta _{2k}}\right]
\approx Z_{\alpha _{2k}}^{\alpha _{2k-1}}\omega _{\alpha _{2k-1}\beta
_{2k-1}}Z_{\beta _{2k}}^{\beta _{2k-1}}+A_{\alpha _{2k}}^{\alpha
_{2k+1}}\omega _{\alpha _{2k+1}\beta _{2k+1}}A_{\beta _{2k}}^{\beta _{2k+1}},
\label{l36} \\
&&\left[ \tilde{\chi}_{\alpha _{2k-2}},\tilde{\chi}_{\beta _{2k}}\right]
\approx A_{\alpha _{2k-2}}^{\alpha _{2k-1}}\omega _{\alpha _{2k-1}\beta
_{2k-1}}Z_{\beta _{2k}}^{\beta _{2k-1}},  \label{l37}
\end{eqnarray}%
with $k=\overline{1,\left[ \frac{L}{2}\right] }$;

\noindent -if $L$ even
\begin{eqnarray}
&&\left[ \tilde{\chi}_{\alpha _{0}},\tilde{\chi}_{\beta _{0}}\right] \approx
\mu _{\alpha _{0}\beta _{0}}^{\left( L\right) },  \label{l38} \\
&&\left[ \tilde{\chi}_{\alpha _{2k}},\tilde{\chi}_{\beta _{2k}}\right]
\approx Z_{\alpha _{2k}}^{\alpha _{2k-1}}\omega _{\alpha _{2k-1}\beta
_{2k-1}}Z_{\beta _{2k}}^{\beta _{2k-1}}+A_{\alpha _{2k}}^{\alpha
_{2k+1}}\omega _{\alpha _{2k+1}\beta _{2k+1}}A_{\beta _{2k}}^{\beta _{2k+1}},
\label{l39} \\
&&\left[ \tilde{\chi}_{\alpha _{2k-2}},\tilde{\chi}_{\beta _{2k}}\right]
\approx A_{\alpha _{2k-2}}^{\alpha _{2k-1}}\omega _{\alpha _{2k-1}\beta
_{2k-1}}Z_{\beta _{2k}}^{\beta _{2k-1}},  \label{l40} \\
&&\left[ \tilde{\chi}_{\alpha _{L}},\tilde{\chi}_{\beta _{L}}\right] \approx
Z_{\alpha _{L}}^{\alpha _{L-1}}\omega _{\alpha _{L-1}\beta _{L-1}}Z_{\beta
_{L}}^{\beta _{L-1}},  \label{l41}
\end{eqnarray}%
with $k=\overline{1,\frac{L}{2}-1}$ in (\ref{l39}) and $k=\overline{1,\frac{L%
}{2}}$ in (\ref{l40}).

Accordingly, the matrix of elements given in (\ref{l24}) reads as
\begin{equation}
C_{\Delta \Delta ^{\prime }}=\left(
\begin{array}{cccc}
\mu _{\alpha _{0}\beta _{0}}^{\left( L\right) } & A_{\alpha _{0}}^{\alpha
_{1}}\omega _{\alpha _{1}\beta _{1}}Z_{\beta _{2}}^{\beta _{1}} & \mathbf{0}
&  \\
Z_{\alpha _{2}}^{\alpha _{1}}\omega _{\alpha _{1}\beta _{1}}A_{\beta
_{0}}^{\beta _{1}} & \phi _{\alpha _{2}\beta _{2}} & A_{\alpha _{2}}^{\alpha
_{3}}\omega _{\alpha _{3}\beta _{3}}Z_{\beta _{4}}^{\beta _{3}} &  \\
\mathbf{0} & Z_{\alpha _{4}}^{\alpha _{3}}\omega _{\alpha _{3}\beta
_{3}}A_{\beta _{2}}^{\beta _{3}} & \phi _{\alpha _{4}\beta _{4}} &  \\
&  &  & \ddots%
\end{array}%
\right) ,  \label{l42}
\end{equation}%
where%
\begin{equation}
\phi _{\alpha _{2k}\beta _{2k}}=Z_{\alpha _{2k}}^{\alpha _{2k-1}}\omega
_{\alpha _{2k-1}\beta _{2k-1}}Z_{\beta _{2k}}^{\beta _{2k-1}}+A_{\alpha
_{2k}}^{\alpha _{2k+1}}\omega _{\alpha _{2k+1}\beta _{2k+1}}A_{\beta
_{2k}}^{\beta _{2k+1}}.  \label{l43}
\end{equation}%
The last block on the main diagonal of (\ref{l42}) is of the type (\ref{l43}%
), with $k=\left[ \frac{L}{2}\right] $ for $L$ odd or respectively of the
form%
\begin{equation}
\phi _{\alpha _{L}\beta _{L}}=Z_{\alpha _{L}}^{\alpha _{L-1}}\omega _{\alpha
_{L-1}\beta _{L-1}}Z_{\beta _{L}}^{\beta _{L-1}}  \label{l44}
\end{equation}%
for $L$ even. The invertibility of $C_{\Delta \Delta ^{\prime }}$ is
obtained by constructing its inverse, which can be checked to have the
expression
\begin{equation}
C^{\Delta ^{\prime }\Delta ^{\prime \prime }}=\left(
\begin{array}{cccc}
\mu ^{\left( L\right) \beta _{0}\rho _{0}} & Z_{\gamma _{1}}^{\beta _{0}}%
\hat{e}_{\sigma _{1}}^{\gamma _{1}}\omega ^{\sigma _{1}\lambda _{1}}\bar{A}%
_{\lambda _{1}}^{\rho _{2}} & \mathbf{0} &  \\
\bar{A}_{\sigma _{1}}^{\beta _{2}}\omega ^{\sigma _{1}\lambda _{1}}\hat{e}%
_{\lambda _{1}}^{\gamma _{1}}Z_{\gamma _{1}}^{\rho _{0}} & \psi ^{\beta
_{2}\rho _{2}} & Z_{\tau _{3}}^{\beta _{2}}\hat{e}_{\gamma _{3}}^{\tau
_{3}}\omega ^{\gamma _{3}\lambda _{3}}\bar{A}_{\lambda _{3}}^{\rho _{4}} &
\\
\mathbf{0} & \bar{A}_{\lambda _{3}}^{\beta _{4}}\omega ^{\lambda _{3}\gamma
_{3}}\hat{e}_{\gamma _{3}}^{\tau _{3}}Z_{\tau _{3}}^{\rho _{2}} & \psi
^{\beta _{4}\rho _{4}} &  \\
&  &  & \ddots%
\end{array}%
\right) ,  \label{l45}
\end{equation}%
with
\begin{equation}
\psi ^{\beta _{2k}\rho _{2k}}=\bar{A}_{\sigma _{2k-1}}^{\beta _{2k}}\omega
^{\sigma _{2k-1}\lambda _{2k-1}}\bar{A}_{\lambda _{2k-1}}^{\rho
_{2k}}+Z_{\tau _{2k+1}}^{\beta _{2k}}\hat{e}_{\gamma _{2k+1}}^{\tau
_{2k+1}}\omega ^{\gamma _{2k+1}\lambda _{2k+1}}\hat{e}_{\lambda
_{2k+1}}^{\sigma _{2k+1}}Z_{\sigma _{2k+1}}^{\rho _{2k}}.  \label{l46}
\end{equation}%
The last block on the main diagonal of (\ref{l45}) is given by (\ref{l46}),
with $k=\left[ \frac{L}{2}\right] $ for $L$ odd or respectively
\begin{equation}
\psi ^{\beta _{L}\rho _{L}}=\bar{A}_{\sigma _{L-1}}^{\beta _{L}}\omega
^{\sigma _{L-1}\lambda _{L-1}}\bar{A}_{\lambda _{L-1}}^{\rho _{L}}
\label{l48}
\end{equation}%
for $L$ even. Indeed, simple computation yields
\begin{equation}
C_{\Delta \Delta ^{\prime }}C^{\Delta ^{\prime }\Delta ^{\prime \prime
}}\approx \left(
\begin{array}{cccc}
\delta _{\alpha _{0}}^{\rho _{0}} & \mathbf{0} & \mathbf{0} &  \\
\mathbf{0} & \delta _{\alpha _{2}}^{\rho _{2}} & \mathbf{0} &  \\
\mathbf{0} & \mathbf{0} & \delta _{\alpha _{4}}^{\rho _{_{4}}} &  \\
&  &  & \ddots%
\end{array}%
\right) ,  \label{l50}
\end{equation}%
such that (\ref{l42}) is indeed invertible and its inverse is expressed by (%
\ref{l45}). This proves (ii).

(iii) As (\ref{l42}) is invertible, it follows that it displays no null
vectors and hence the functions $\tilde{\chi}_{\Delta }$ are all independent
or, in other words, the constraint set (\ref{l14}) and (\ref{l15}) (or (\ref%
{l16})--(\ref{l18})) is irreducible. This proves (iii). $\Box $

Taking into account the result given by (\ref{l45}), it follows that the
Dirac bracket built with respect to the irreducible second-class constraints
(\ref{l14}) and (\ref{l15}) (or (\ref{l16})--(\ref{l18}))
\begin{equation}
\left. \left[ F,G\right] ^{\left( L\right) \ast }\right\vert _{\mathrm{ired}%
}=\left[ F,G\right] -\left[ F,\tilde{\chi}_{\Delta }\right] C^{\Delta \Delta
^{\prime }}\left[ \tilde{\chi}_{\Delta ^{\prime }},G\right]  \label{l51}
\end{equation}%
takes the particular form
\begin{eqnarray}
&&\left. \left[ F,G\right] ^{\left( L\right) \ast }\right\vert _{\mathrm{ired%
}}=\left[ F,G\right] -\left[ F,\tilde{\chi}_{\alpha _{0}}\right] \mu
^{\left( L\right) \alpha _{0}\beta _{0}}\left[ \tilde{\chi}_{\beta _{0}},G%
\right]  \notag \\
&&-\sum\limits_{k=0}^{\left[ \frac{L}{2}\right] -1}\left\{ \left[ F,\tilde{%
\chi}_{\alpha _{2k}}\right] Z_{\alpha _{2k+1}}^{\alpha _{2k}}\hat{e}_{\gamma
_{2k+1}}^{\alpha _{2k+1}}\omega ^{\gamma _{2k+1}\beta _{2k+1}}\bar{A}_{\beta
_{2k+1}}^{\beta _{2k+2}}\left[ \tilde{\chi}_{\beta _{2k+2}},G\right] \right.
\notag \\
&&+\left[ F,\tilde{\chi}_{\alpha _{2k+2}}\right] \bar{A}_{\alpha
_{2k+1}}^{\alpha _{2k+2}}\omega ^{\alpha _{2k+1}\gamma _{2k+1}}\hat{e}%
_{\gamma _{2k+1}}^{\beta _{2k+1}}Z_{\beta _{2k+1}}^{\beta _{2k}}\left[
\tilde{\chi}_{\beta _{2k}},G\right]  \notag \\
&&\left. +\left[ F,\tilde{\chi}_{\alpha _{2k+2}}\right] \psi ^{\alpha
_{2k+2}\beta _{2k+2}}\left[ \tilde{\chi}_{\beta _{2k+2}},G\right] \right\} .
\label{l52}
\end{eqnarray}

\begin{theorem}
\label{th9} The Dirac bracket with respect to the irreducible second-class
constraints (\ref{l52}) coincides with that of the intermediate system
\begin{equation}
\left. \left[ F,G\right] ^{\left( L\right) \ast }\right\vert _{\mathrm{ired}%
}\approx \left. \left[ F,G\right] ^{\left( L\right) \ast }\right\vert _{z,y}.
\label{l53}
\end{equation}
\end{theorem}

\textbf{Proof. }We start from the right-hand side of (\ref{l52}) and show
that it is (weakly) equal to the right-hand side of (\ref{l10}). By direct
computation, we obtain that:%
\begin{eqnarray}
&&\left[ F,\tilde{\chi}_{\alpha _{0}}\right] \mu ^{\left( L\right) \alpha
_{0}\beta _{0}}\left[ \tilde{\chi}_{\beta _{0}},G\right] \approx  \notag \\
&&\left[ F,\chi _{\alpha _{0}}\right] \mu ^{\left( L\right) \alpha _{0}\beta
_{0}}\left[ \chi _{\beta _{0}},G\right] +\left[ F,y_{\alpha _{1}}\right]
D_{\sigma _{1}}^{\alpha _{1}}\omega ^{\sigma _{1}\lambda _{1}}D_{\lambda
_{1}}^{\beta _{1}}\left[ y_{\beta _{1}},G\right] ,  \label{l54} \\
&&\left[ F,\tilde{\chi}_{\alpha _{2k}}\right] Z_{\alpha _{2k+1}}^{\alpha
_{2k}}\hat{e}_{\gamma _{2k+1}}^{\alpha _{2k+1}}\omega ^{\gamma _{2k+1}\beta
_{2k+1}}\bar{A}_{\beta _{2k+1}}^{\beta _{2k+2}}\left[ \tilde{\chi}_{\beta
_{2k+2}},G\right] \approx  \notag \\
&&\left[ F,y_{\alpha _{2k+1}}\right] D_{\gamma _{2k+1}}^{\alpha
_{2k+1}}\omega ^{\gamma _{2k+1}\lambda _{2k+1}}\left( \delta _{\lambda
_{2k+1}}^{\beta _{2k+1}}-D_{\lambda _{2k+1}}^{\beta _{2k+1}}\right) \left[
y_{\beta _{2k+1}},G\right] , \\
&&\left[ F,\tilde{\chi}_{\alpha _{2k+2}}\right] \bar{A}_{\alpha
_{2k+1}}^{\alpha _{2k+2}}\omega ^{\alpha _{2k+1}\gamma _{2k+1}}\hat{e}%
_{\gamma _{2k+1}}^{\beta _{2k+1}}Z_{\beta _{2k+1}}^{\beta _{2k}}\left[
\tilde{\chi}_{\beta _{2k}},G\right] \approx  \notag \\
&&\left[ F,y_{\alpha _{2k+1}}\right] \left( \delta _{\gamma _{2k+1}}^{\alpha
_{2k+1}}-D_{\gamma _{2k+1}}^{\alpha _{2k+1}}\right) \omega ^{\gamma
_{2k+1}\lambda _{2k+1}}D_{\lambda _{2k+1}}^{\beta _{2k+1}}\left[ y_{\beta
_{2k+1}},G\right] ,
\end{eqnarray}%
with $k=\overline{0,\left[ \frac{L}{2}\right] -1}$. Also direct computation
provides:

\noindent -if $L$ odd%
\begin{eqnarray}
&&\left[ F,\tilde{\chi}_{\alpha _{2k+2}}\right] \psi ^{\alpha _{2k+2}\beta
_{2k+2}}\left[ \tilde{\chi}_{\beta _{2k+2}},G\right] \approx  \notag \\
&&\left[ F,y_{\alpha _{2k+1}}\right] \left( \delta _{\gamma _{2k+1}}^{\alpha
_{2k+1}}-D_{\gamma _{2k+1}}^{\alpha _{2k+1}}\right) \omega ^{\gamma
_{2k+1}\lambda _{2k+1}}\left( \delta _{\lambda _{2k+1}}^{\beta
_{2k+1}}-D_{\lambda _{2k+1}}^{\beta _{2k+1}}\right) \left[ y_{\beta
_{2k+1}},G\right]  \notag \\
&&+\left[ F,y_{\alpha _{2k+3}}\right] D_{\gamma _{2k+3}}^{\alpha
_{2k+3}}\omega ^{\gamma _{2k+3}\lambda _{2k+3}}D_{\lambda _{2k+3}}^{\beta
_{2k+3}}\left[ y_{\beta _{2k+3}},G\right] ,
\end{eqnarray}%
with $k=\overline{0,\left[ \frac{L}{2}\right] -1}$;

\noindent -if $L$\ even%
\begin{eqnarray}
&&\left[ F,\tilde{\chi}_{\alpha _{2k+2}}\right] \psi ^{\alpha _{2k+2}\beta
_{2k+2}}\left[ \tilde{\chi}_{\beta _{2k+2}},G\right] \approx  \notag \\
&&\left[ F,y_{\alpha _{2k+1}}\right] \left( \delta _{\gamma _{2k+1}}^{\alpha
_{2k+1}}-D_{\gamma _{2k+1}}^{\alpha _{2k+1}}\right) \omega ^{\gamma
_{2k+1}\lambda _{2k+1}}\left( \delta _{\lambda _{2k+1}}^{\beta
_{2k+1}}-D_{\lambda _{2k+1}}^{\beta _{2k+1}}\right) \left[ y_{\beta
_{2k+1}},G\right]  \notag \\
&&+\left[ F,y_{\alpha _{2k+3}}\right] D_{\gamma _{2k+3}}^{\alpha
_{2k+3}}\omega ^{\gamma _{2k+3}\lambda _{2k+3}}D_{\lambda _{2k+3}}^{\beta
_{2k+3}}\left[ y_{\beta _{2k+3}},G\right] ,
\end{eqnarray}%
with $k=\overline{0,\frac{L}{2}-2}$.

Further computation finally gives:
\begin{eqnarray}
&\left[ F,\tilde{\chi}_{\alpha _{L}}\right] \bar{A}_{\gamma _{L-1}}^{\alpha
_{L}}\omega ^{\gamma _{L-1}\lambda _{L-1}}\bar{A}_{\lambda _{L-1}}^{\beta
_{L}}\left[ \tilde{\chi}_{\beta _{L}},G\right] \approx  \notag \\
&\left[ F,y_{\alpha _{L-1}}\right] \left( \delta _{\gamma _{L-1}}^{\alpha
_{L-1}}-D_{\gamma _{L-1}}^{\alpha _{L-1}}\right) \omega ^{\gamma
_{L-1}\lambda _{L-1}}\left( \delta _{\lambda _{L-1}}^{\beta
_{L-1}}-D_{\lambda _{L-1}}^{\beta _{L-1}}\right) \left[ y_{\beta _{L-1}},G%
\right] .
\end{eqnarray}

Inserting the last formulae in (\ref{l52}) we arrive at (\ref{l53}), which
proves the theorem. $\Box $

\subsection{Main result}

Based on (\ref{l13}) and (\ref{l53}), we are led to the relation%
\begin{equation}
\left[ F,G\right] ^{\left( L\right) \ast }\approx \left. \left[ F,G\right]
^{\left( L\right) \ast }\right\vert _{\mathrm{ired}},  \label{u76}
\end{equation}%
which expresses the fact that second-class constraints reducible of an
arbitrary order $L$ can be systematically approached in an irreducible
manner. This is the key result of the present paper.

\subsection{Geometrical interpretation of the irreducible approach}

Let us denote by $P$\ the original phase-space and by $P^{\prime }$ the
phase-space of the intermediate system, and hence also of the irreducible
theory. Both are symplectic manifolds endowed with symplectic two-forms
whose coefficients are in each case the elements of the inverse of the
matrix having as elements the fundamental Poisson brackets. We denote by $%
\Sigma $\ and respectively $\Sigma ^{\prime }$\ the second-class constraint
surface for the original system and respectively for the intermediate
theory. By Theorem \ref{th8} it follows that the second-class constraint
surface of the irreducible system, given by equations (\ref{l14}) and (\ref%
{l15}) (or (\ref{l16})--(\ref{l18})), is nothing but an equivalent
representation of $\Sigma ^{\prime }$. Let $j$ and respectively $j^{\prime }$
be the injective immersions of $\Sigma $\ in $P$\ and respectively of $%
\Sigma ^{\prime }$ in $P^{\prime }$. The second-class property of $\Sigma $\
and respectively of $\Sigma ^{\prime }$\ is equivalent to the fact that the
induced symplectic two-forms $j^{\ast }\omega $\ and respectively $%
j^{^{\prime }\ast }\omega ^{\prime }$\ are non-degenerate \cite{7}, which is
the same with \cite{sundermeyer}
\begin{equation}
j_{\ast }\left( T\Sigma \right) \cap T\Sigma ^{\perp }=\left\{ 0\right\}
,\quad j_{\ast }^{\prime }\left( T\Sigma ^{\prime }\right) \cap T\Sigma
^{\prime \perp }=\left\{ 0\right\} .  \label{rnewg1}
\end{equation}

It is easy to argue now the preservation of the original number of physical
degrees of freedom with respect to the intermediate and irreducible systems.
The dimensions of the original and respectively of the intermediate or
irreducible phase-space are valued as
\begin{equation}
\dim P=2N,\qquad \dim P^{\prime }=2N+\sum\limits_{k=0}^{\left[ \frac{L-1}{2}%
\right] }M_{2k+1},  \label{rnewg2}
\end{equation}%
while the dimensions of the corresponding submanifolds $\Sigma $\ and
respectively $\Sigma ^{\prime }$\ are equal by construction
\begin{equation}
2\mathcal{N}=\dim \Sigma =\dim \Sigma ^{\prime
}=2N-\sum\limits_{l=0}^{L}\left( -\right) ^{l}M_{l}.  \label{rnewg3}
\end{equation}%
Because the induced symplectic two-forms $j^{\ast }\omega $\ and
respectively $j^{^{\prime }\ast }\omega ^{\prime }$\ are non-degenerate,
from (\ref{rnewg3}) we deduce that%
\begin{equation}
\mathrm{rank}\left( j^{\ast }\omega \right) =\mathrm{rank}\left( j^{\prime
\ast }\omega ^{\prime }\right) =2\mathcal{N},  \label{rnewg4}
\end{equation}%
and therefore all the three systems, original, intermediate, and
irreducible, possess the same number of physical degrees of freedom, $N$,
defined as half of the rank of the induced two-forms.

Moreover, the induced symplectic two-forms $j^{\ast }\omega $\ and $%
j^{^{\prime }\ast }\omega ^{\prime }$\ can be brought to exactly the same
form in some conveniently chosen charts. For instance, if we (locally)
parameterize the submanifolds $\Sigma $\ and $\Sigma ^{\prime }$\ (having
the same dimension $2N$)\ by the coordinates $\left( \xi ^{\alpha }\right)
_{\alpha =\overline{1,2\mathcal{N}}}$, then the local expressions of the
immersions $j$\ and respectively $j^{\prime }$\ read as%
\begin{equation}
z^{a}=z^{a}\left( \xi \right) ,\qquad a=\overline{1,2N}  \label{rnewg5}
\end{equation}%
and respectively%
\begin{equation}
\left\{
\begin{array}{c}
z^{a}=z^{a}\left( \xi \right) ,\qquad a=\overline{1,2N}, \\
y^{A_{2k+1}}=0,\qquad A_{2k+1}=\overline{1,M_{2k+1}},\qquad k=\overline{0,%
\left[ \frac{L-1}{2}\right] }.%
\end{array}%
\right.   \label{rnewg6}
\end{equation}%
Obviously, related to the local expressions of (\ref{rnewg5}) and (\ref%
{rnewg6}) we have that%
\begin{equation}
\left( j^{\ast }\omega \right) _{\alpha \beta }=\left( j^{\prime \ast
}\omega ^{\prime }\right) _{\alpha \beta },\qquad \forall \alpha ,\beta =%
\overline{1,2\mathcal{N}}.  \label{rnewg7}
\end{equation}

One of the main benefits enabled by our irreducible construction is the
computation in a standard manner of the coefficients of the induced
symplectic two-form (\ref{rnewg7}) as the elements of the inverse of the
matrix having as elements the fundamental Dirac brackets (see Theorem 2.5
from \cite{7}). By `standard' we mean without need to take any specific
parametrization of the second-class constraint surface and, implicitly, to
perform any separation into dependent and independent constraint functions.

We have seen that the matrices $D_{\alpha _{k}}^{\gamma _{k}}$\ (with $k>0$)
are some intermediate steps required by the irreducible procedure, which
serve to the construction of the projection $D_{\alpha _{0}}^{\gamma _{0}}$,
which projects\ the system of local generators%
\begin{equation}
X_{\alpha _{0}}=\sigma ^{ab}\frac{\partial \chi _{\alpha _{0}}}{\partial
z^{a}}\frac{\partial }{\partial z^{b}}  \label{rnewg8}
\end{equation}%
of the space $T\Sigma ^{\perp }$\ into a local basis of the same space.

\section{Example\label{pforms}}

Let us exemplify the general theory on gauge-fixed Abelian $p$-form gauge
fields. Abelian $p$-forms are described by the Lagrangian action%
\begin{equation}
S_{0}^{\mathrm{L}}\left[ A_{\mu _{1}\cdots \mu _{p}}\right] =\int \mathrm{d}%
^{D}x\left( -\frac{1}{2\cdot \left( p+1\right) !}F_{\mu _{1}\ldots \mu
_{p+1}}F^{\mu _{1}\ldots \mu _{p+1}}\right) ,  \label{mo1}
\end{equation}%
where the field strength of $A_{\mu _{1}\ldots \mu _{p}}$\ is defined in the
standard manner by $F_{\mu _{1}\ldots \mu _{p+1}}\equiv \partial _{\lbrack
\mu _{1}}A_{\mu _{2}\ldots \mu _{p+1}]}$.\ Furthermore, we take the
spacetime dimension $D$\ to satisfy $D\geq p+1$, since otherwise the number
of physical degrees of freedom\ would be strictly negative. Everywhere in
the sequel the notation $\left[ \mu \ldots \nu \right] $\ signifies
antisymmetry with respect to all the indices between brackets without
normalization factors (i.e., the independent terms appear only once and are
not multiplied by overall numerical factors). We will briefly expose the
canonical analysis of Abelian $p$-forms. For more details, see \cite%
{baulieumarc} and \cite{costicaodile}.

From the definitions of canonical momenta\footnote{%
We work with a `mostly negative' metric tensor $\left( +--\ldots -\right) $,
such that no confusion arises in the notation $\dot{A}^{\mu _{1}\cdots \mu
_{p}}$ for the time derivative of $A^{\mu _{1}\cdots \mu _{p}}$. }
\begin{equation}
\pi _{\mu _{1}\cdots \mu _{p}}=\frac{\partial \mathcal{L}_{0}}{\partial \dot{%
A}^{\mu _{1}\cdots \mu _{p}}}  \label{mo2}
\end{equation}%
on the one hand one obtains the primary constraints
\begin{equation}
G_{i_{1}\ldots i_{p-1}}^{(1)}\equiv \pi _{0i_{1}\ldots i_{p-1}}\approx 0,
\label{mo3}
\end{equation}%
and, on the other hand, one expresses the time derivatives of $%
A_{i_{1}\cdots i_{p}}$\ as
\begin{equation}
\dot{A}_{i_{1}\cdots i_{p}}=-p!\pi _{i_{1}\cdots i_{p}}-\left( -\right)
^{p}\partial _{\lbrack i_{1}}A_{i_{2}\cdots i_{p}]0}.  \label{mo4}
\end{equation}%
The canonical Hamiltonian in defined in the standard manner for constrained
systems \cite{7} and reduces to%
\begin{eqnarray}
H &=&\int \mathrm{d}^{D-1}\mathbf{x}\left( -pA^{0i_{1}\ldots
i_{p-1}}\partial ^{l}\pi _{li_{1}\ldots i_{p-1}}\right.   \notag \\
&&\left. \mathbf{-}\frac{p!}{2}\pi _{i_{1}\cdots i_{p}}\pi ^{i_{1}\cdots
i_{p}}+\frac{1}{2\cdot \left( p+1\right) !}F_{i_{1}\cdots
i_{p+1}}F^{i_{1}\cdots i_{p+1}}\right) ,  \label{mo5}
\end{eqnarray}%
where we made the notation $x=\left( x^{0},\mathbf{x}\right) $.

Dirac's algorithm (the consistency conditions for the primary constraints (%
\ref{mo3})) provides the secondary constraints%
\begin{equation}
\chi _{i_{1}\ldots i_{p-1}}^{(1)}\equiv -p\partial ^{l}\pi _{li_{1}\ldots
i_{p-1}}\approx 0  \label{model1}
\end{equation}%
and stops after the first step. Therefore, Abelian $p$-form gauge fields are
subject to the constraints (\ref{mo3}) and (\ref{model1}), which are
first-class and, moreover, Abelian (the Poisson brackets among the
constraint functions vanish strongly). It is easy to check the relations%
\begin{equation}
\left[ H,G_{i_{1}\ldots i_{p-1}}^{(1)}\right] =\chi _{i_{1}\ldots
i_{p-1}}^{(1)},  \label{mo8}
\end{equation}%
\begin{equation}
\left[ H,\chi _{i_{1}\ldots i_{p-1}}^{(1)}\right] =0,  \label{mo7}
\end{equation}%
which show that (\ref{mo5}) is also a first-class Hamiltonian for Abelian $p$%
-form gauge fields. The primary first-class constraints are irreducible,
while the secondary first-class ones are off-shell reducible (meaning that
the null eigenvector equations for the constraint functions and for all the
higher-order reducibility functions hold strongly, everywhere on the
phase-space, and not only on the first-class surface) of order $\left(
p-1\right) $. The associated reducibility functions are given below.

It is known that the first-class constraints produce some local
transformations of the canonical variables, which do not affect the physical
state of the system. They are called Hamiltonian gauge transformations.
Although only the primary first-class constraints can be shown to generate
gauge transformations, we accept Dirac's conjecture, according to which all
first-class constraint generate Hamiltonian gauge transformations. The
dynamics of first-class systems is thus not fixed in the sense that for some
fixed initial set of canonical variables, the solution to the Hamiltonian
equations of motion in the presence of first-class constraints is not
unique. In other words, a given physical state of a first-class system is
expressed by more than one set of canonical variables (any two such sets are
related by a Hamiltonian gauge transformation). In practice, it is useful to
eliminate this ambiguity and restore a one-to-one correspondence between
physical states and values of the independent canonical variables. This is
realized via the so called `gauge-fixing procedure' by means of imposing
further restrictions on the canonical variables, known as `canonical gauge
conditions'. These must be `good' canonical gauge conditions in the sense of
\cite{7}, subsection 1.4.1. It is easy to see that a set of good canonical
gauge conditions with respect to the first-class constraints (\ref{mo3}) and
(\ref{model1}) reads as
\begin{eqnarray}
\bar{\chi}^{(1)i_{1}\ldots i_{p-1}} &\equiv &A^{0i_{1}\ldots i_{p-1}}\approx
0,  \label{mo9a} \\
\chi ^{\left( 2\right) j_{1}\ldots j_{p-1}} &\equiv &-\partial
_{m}A^{mj_{1}\ldots j_{p-1}}\approx 0.  \label{model2}
\end{eqnarray}%
The overall constraint set formed with the first-class constraints (\ref{mo3}%
) and (\ref{model1}) together with the chosen canonical gauge conditions (%
\ref{mo9a}) and (\ref{model2}) is a second-class constraint set, off-shell
reducible of order $\left( p-1\right) $. In fact, only (\ref{model1}) and (%
\ref{model2}) are reducible, each of order $\left( p-1\right) $, while (\ref%
{mo3}) and (\ref{mo9a}) are irreducible.

Due to the fact that the second-class constraints (\ref{mo3}) and (\ref{mo9a}%
) are independent, we will eliminate them from the theory by means of the
Dirac bracket built with respect to themselves and will treat along the
irreducible approach exposed in the main body of this paper only the
reducible second-class constraints (\ref{model1}) and (\ref{model2}) It is
useful to organize these second-class constraints in a column vector
\begin{equation}
\chi _{\alpha _{0}}=\left(
\begin{array}{c}
\chi _{i_{1}\ldots i_{p-1}}^{\left( 1\right) } \\
\chi ^{\left( 2\right) j_{1}\ldots j_{p-1}}%
\end{array}%
\right) \approx 0.  \label{l55}
\end{equation}%
Constraints (\ref{l55}) are $\left( p-1\right) $-order reducible, with the
reducibility functions of the form%
\begin{equation}
Z_{\alpha _{k+1}}^{\alpha _{k}}=\left(
\begin{array}{cc}
\frac{1}{\left( p-k-2\right) !}\delta _{m_{1}}^{[i_{1}}\ldots \delta
_{m_{p-k-2}}^{i_{p-k-2}}\partial ^{i_{p-k-1}]} & \mathbf{0} \\
\mathbf{0} & \frac{1}{\left( p-k-1\right) !}\delta _{\lbrack
j_{1}}^{n_{1}}\ldots \delta _{j_{p-k-2}}^{n_{p-k-2}}\partial _{j_{p-k-1}]}%
\end{array}%
\right) ,  \label{l56}
\end{equation}%
for $k=\overline{0,p-2}$. The matrix of the Poisson brackets among the
constraint functions from (\ref{l55}) is expressed by%
\begin{equation}
C_{\alpha _{0}\beta _{0}}=\left(
\begin{array}{cc}
\mathbf{0} & \Delta D_{i_{1}\ldots i_{p-1}}^{j_{1}\ldots j_{p-1}} \\
-\Delta D_{i_{1}\ldots i_{p-1}}^{j_{1}\ldots j_{p-1}} & \mathbf{0}%
\end{array}%
\right) ,  \label{l57}
\end{equation}%
where%
\begin{equation}
D_{i_{1}\ldots i_{p-1}}^{j_{1}\ldots j_{p-1}}=\frac{1}{\left( p-1\right) !}%
\left( \delta _{\lbrack i_{1}}^{j_{1}}\ldots \delta _{i_{p-1}]}^{j_{p-1}}-%
\frac{\delta _{\lbrack i_{1}}^{m_{1}}\ldots \delta
_{i_{p-2}}^{m_{p-2}}\partial _{i_{p-1}]}\delta _{m_{1}}^{[j_{1}}\ldots
\delta _{m_{p-2}}^{j_{p-2}}\partial ^{j_{p-1}]}}{\left( p-2\right) !\Delta }%
\right)   \label{l58}
\end{equation}%
and $\Delta =\partial _{i}\partial ^{i}$.

In this particular case the functions $\left( \bar{A}_{\alpha _{k}}^{\alpha
_{k+1}}\right) _{k=\overline{0,p-2}}$ read as%
\begin{equation}
\bar{A}_{\alpha _{k}}^{\alpha _{k+1}}=\left(
\begin{array}{cc}
\frac{1}{\left( p-k-1\right) !\Delta }\delta _{\lbrack j_{1}}^{m_{1}}\ldots
\delta _{j_{p-k-2}}^{m_{p-k-2}}\partial _{j_{p-k-1}]} & \mathbf{0} \\
\mathbf{0} & \frac{1}{\left( p-k-2\right) !\Delta }\delta
_{n_{1}}^{[i_{1}}\ldots \delta _{n_{p-k-2}}^{i_{p-k-2}}\partial ^{i_{p-k-1}]}%
\end{array}%
\right) .  \label{l59}
\end{equation}%
If we take $\bar{A}_{\alpha _{0}}^{\alpha _{1}}$ as in (\ref{l59}) for $k=0$%
, then we find that $D_{\alpha _{0}}^{\beta _{0}}$ is given by%
\begin{equation}
D_{\alpha _{0}}^{\beta _{0}}=\left(
\begin{array}{cc}
D_{i_{1}\ldots i_{p-1}}^{j_{1}\ldots j_{p-1}} & \mathbf{0} \\
\mathbf{0} & D_{i_{1}\ldots i_{p-1}}^{j_{1}\ldots j_{p-1}}%
\end{array}%
\right) .  \label{l60}
\end{equation}%
From equation (\ref{l5}) we obtain $M^{\left( p-1\right) \alpha _{0}\beta
_{0}}$ under the form
\begin{equation}
M^{\left( p-1\right) \alpha _{0}\beta _{0}}=\left(
\begin{array}{cc}
\mathbf{0} & -\frac{1}{\Delta }D_{j_{1}\ldots j_{p-1}}^{i_{1}\ldots i_{p-1}}
\\
\frac{1}{\Delta }D_{j_{1}\ldots j_{p-1}}^{i_{1}\ldots i_{p-1}} & \mathbf{0}%
\end{array}%
\right) .  \label{l61}
\end{equation}%
With $M^{\left( p-1\right) \alpha _{0}\beta _{0}}$ at hand, we are able to
construct the Dirac bracket given by (\ref{l6}). After some computation, we
determine the fundamental Dirac brackets as:%
\begin{eqnarray}
&&\left[ A^{i_{1}\ldots i_{p}}\left( x\right) ,\pi _{j_{1}\ldots
j_{p}}\left( y\right) \right] _{x_{0}=y_{0}}^{\ast }=D_{j_{1}\ldots
j_{p}}^{i_{1}\ldots i_{p}}\delta ^{D-1}\left( \mathbf{x}-\mathbf{y}\right) ,
\label{l62} \\
&&\left[ A^{i_{1}\ldots i_{p}}\left( x\right) ,A^{j_{1}\ldots j_{p}}\left(
y\right) \right] _{x_{0}=y_{0}}^{\ast }=0=\left[ \pi _{i_{1}\ldots
i_{p}}\left( x\right) ,\pi _{j_{1}\ldots j_{p}}\left( y\right) \right]
_{x_{0}=y_{0}}^{\ast },  \label{l621}
\end{eqnarray}%
where
\begin{equation}
D_{i_{1}\ldots i_{p}}^{j_{1}\ldots j_{p}}=\frac{1}{p!}\left( \delta
_{\lbrack i_{1}}^{j_{1}}\ldots \delta _{i_{p}]}^{j_{p}}-\frac{\delta
_{\lbrack i_{1}}^{m_{1}}\ldots \delta _{i_{p-1}}^{m_{p-1}}\partial
_{i_{p}]}\delta _{m_{1}}^{[j_{1}}\ldots \delta _{m_{p-1}}^{j_{p-1}}\partial
^{j_{p}]}}{\left( p-1\right) !\Delta }\right) .  \label{l63}
\end{equation}%
Formula (\ref{l71}) together with (\ref{l60}) and (\ref{l61}) provides
\begin{equation}
\mu ^{\left( p-1\right) \alpha _{0}\beta _{0}}=\left(
\begin{array}{cc}
\mathbf{0} & -\frac{1}{\left( p-1\right) !\Delta }\delta _{\lbrack
j_{1}}^{i_{1}}\ldots \delta _{j_{p-1}]}^{i_{p-1}} \\
\frac{1}{\left( p-1\right) !\Delta }\delta _{\lbrack i_{1}}^{j_{1}}\ldots
\delta _{i_{p-1}]}^{j_{p-1}} & \mathbf{0}%
\end{array}%
\right) ,  \label{l64}
\end{equation}%
which clearly exhibits the invertibility of $\mu ^{\left( p-1\right) \alpha
_{0}\beta _{0}}$. By computing the fundamental Dirac brackets with the help
of (\ref{l7}) (with $\mu ^{\left( p-1\right) \alpha _{0}\beta _{0}}$ given
by (\ref{l64})), we reobtain precisely (\ref{l62}) and (\ref{l621}).

In order to construct the irreducible system of second-class constraints
that is equivalent to the original one (like in subsection \ref{constrIIired}%
), we need to enlarge the phase-space by the independent variables $\left(
y_{\alpha _{2k+1}}\right) _{k=\overline{0,\left[ \frac{p}{2}\right] -1}}$
and to know the exact form of the functions $\left( A_{\alpha _{k}}^{\alpha
_{k+1}}\right) _{k=\overline{0,\left[ \frac{p}{2}\right] -1}}$.\ For
gauge-fixed $p$-forms, it is necessary to add the supplementary variables%
\begin{equation}
y_{\alpha _{2k+1}}=\left(
\begin{array}{c}
P_{i_{1}\ldots i_{p-2k-2}} \\
B^{i_{1}\ldots i_{p-2k-2}}%
\end{array}%
\right) ,  \label{l65}
\end{equation}%
with the Poisson brackets%
\begin{equation}
\omega _{\alpha _{2k+1}\beta _{2k+1}}=\left(
\begin{array}{cc}
\mathbf{0} & -\frac{1}{\left( p-2k-2\right) !}\delta _{\lbrack
i_{1}}^{j_{1}}\ldots \delta _{i_{p-2k-2}]}^{j_{p-2k-2}} \\
\frac{1}{\left( p-2k-2\right) !}\delta _{\lbrack j_{1}}^{i_{1}}\ldots \delta
_{j_{p-2k-2]}}^{i_{p-2k-2}} & \mathbf{0}%
\end{array}%
\right) .  \label{l66}
\end{equation}%
The functions $\left( A_{\alpha _{k}}^{\alpha _{k+1}}\right) _{k=\overline{0,%
\left[ \frac{p}{2}\right] -1}}$ can be chosen for instance of the form
\begin{equation}
A_{\alpha _{2k}}^{\alpha _{2k+1}}=\left(
\begin{array}{cc}
\frac{\left( -\right) ^{2k+1}}{\left( p-2k-1\right) !}\delta _{\lbrack
i_{1}}^{m_{1}}\ldots \delta _{i_{p-2k-2}}^{m_{p-2k-2}}\partial _{i_{p-2k-1}]}
& \mathbf{0} \\
\mathbf{0} & \frac{\left( -\right) ^{2k+1}}{\left( p-2k-2\right) !}\delta
_{n_{1}}^{[j_{1}}\ldots \delta _{n_{p-2k-2}}^{j_{p-2k-2}}\partial
^{j_{p-2k-1}]}%
\end{array}%
\right) .  \label{l67}
\end{equation}%
The link between the function sets $\left( A_{\alpha _{2k}}^{\alpha
_{2k+1}}\right) _{k=\overline{0,\left[ \frac{p}{2}\right] -1}}$ and $\left(
\bar{A}_{\alpha _{2k}}^{\alpha _{2k+1}}\right) _{k=\overline{0,\left[ \frac{p%
}{2}\right] -1}}$ is expressed in the case of the model under study by:

\noindent -if $p$ is odd, by relations (\ref{l21}), with $\left( \hat{e}%
_{\alpha _{2k+1}}^{\beta _{2k+1}}\right) _{k=\overline{0,\left[ \frac{p}{2}%
\right] -1}}$ taken as%
\begin{equation}
\hat{e}_{\alpha _{2k+1}}^{\beta _{2k+1}}=\left(
\begin{array}{cc}
\frac{\left( -\right) ^{2k+1}}{\left( p-2k-2\right) !\Delta }\delta
_{\lbrack i_{1}}^{m_{1}}\ldots \delta _{i_{p-2k-2}]}^{m_{p-2k-2}} & \mathbf{0%
} \\
\mathbf{0} & \frac{\left( -\right) ^{2k+1}}{\left( p-2k-2\right) !\Delta }%
\delta _{n_{1}}^{[j_{1}}\ldots \delta _{n_{p-2k-2}}^{j_{p-2k-2}]}%
\end{array}%
\right) ;  \label{l68}
\end{equation}%
\noindent -if $p$ is even, by relations (\ref{l19}) and (\ref{l20}), with $%
\left( \hat{e}_{\alpha _{2k+1}}^{\beta _{2k+1}}\right) _{k=\overline{0,\frac{%
p}{2}-2}}$ given by (\ref{l68}) and $\bar{D}_{\beta _{p-1}}^{\alpha _{p-1}}$
of the form%
\begin{equation}
\bar{D}_{\beta _{p-1}}^{\alpha _{p-1}}=\left(
\begin{array}{cc}
\frac{1}{\Delta } & 0 \\
0 & \frac{1}{\Delta }%
\end{array}%
\right) .  \label{l69}
\end{equation}

The set of irreducible second-class constraints equivalent with (\ref{model1}%
) and (\ref{model2}) follows from formulae (\ref{l16})--(\ref{l18}) for $p$
odd or (\ref{l14}) and (\ref{l15}) for $p$ even and is given by%
\begin{eqnarray}
\tilde{\chi}_{i_{1}\ldots i_{p-1}}^{\left( 1\right) } &\equiv &-p\partial
^{l}\pi _{li_{1}\ldots i_{p-1}}+\frac{\left( -\right) ^{p-1}}{p-1}\partial
_{\lbrack i_{1}}P_{i_{2}\ldots i_{p-1}]}\approx 0, \\
\tilde{\chi}^{\left( 2\right) j_{1}\ldots j_{p-1}} &\equiv &-\partial
_{m}A^{mj_{1}\ldots j_{p-1}}+\left( -\right) ^{p-1}\partial ^{\lbrack
j_{1}}B^{j_{2}\ldots j_{p-1}]}\approx 0,
\end{eqnarray}%
together with:

\noindent -if $p$ odd
\begin{eqnarray}
&&\tilde{\chi}_{i_{1}\ldots i_{p-2k-1}}^{\left( 1\right) }\equiv \left(
-\right) ^{p-1}\left( p-2k\right) \partial ^{l}P_{li_{1}\ldots i_{p-2k-1}}+%
\frac{\left( -\right) ^{p-1}}{p-2k-1}\partial _{\lbrack i_{1}}P_{i_{2}\ldots
i_{p-2k-1}]}, \\
&&\tilde{\chi}^{\left( 2\right) j_{1}\ldots j_{p-2k-1}}\equiv \left(
-\right) ^{p-1}\partial _{m}B^{mj_{1}\ldots j_{p-2k-1}}+\left( -\right)
^{p-1}\partial ^{\lbrack j_{1}}B^{j_{2}\ldots j_{p-2k-1}]}, \\
&&\tilde{\chi}^{\left( 1\right) }\equiv \left( -\right) ^{p-1}\partial
^{l}P_{l}, \\
&&\chi ^{\left( 2\right) }\equiv \left( -\right) ^{p-1}\partial _{m}B^{m},
\end{eqnarray}%
with $k=\overline{1,\left[ \frac{p}{2}\right] -1}$;

\noindent -if $p$ even
\begin{eqnarray}
&&\tilde{\chi}_{i_{1}\ldots i_{p-2k-1}}^{\left( 1\right) }\equiv \left(
-\right) ^{p-1}\left( p-2k\right) \partial ^{l}P_{li_{1}\ldots i_{p-2k-1}}+%
\frac{\left( -\right) ^{p-1}}{p-2k-1}\partial _{\lbrack i_{1}}P_{i_{2}\ldots
i_{p-2k-1}]}, \\
&&\chi ^{\left( 2\right) j_{1}\ldots j_{p-2k-1}}\equiv \left( -\right)
^{p-1}\partial _{m}B^{mj_{1}\ldots j_{p-2k-1}}+\left( -\right)
^{p-1}\partial ^{\lbrack j_{1}}B^{j_{2}\ldots j_{p-2k-1}]},
\end{eqnarray}%
with $k=\overline{1,\frac{p}{2}-1}$.

At this stage, we have constructed all the objects entering the structure of
the irreducible Dirac bracket (\ref{l52}). It is essential to remark that
the irreducible second-class constraints are local. If we construct the
irreducible Dirac bracket and evaluate the fundamental Dirac brackets among
the original variables, then we finally obtain that they are expressed by
relations (\ref{l62}) and (\ref{l621}). This completes the irreducible
analysis of gauge-fixed $p$-form gauge fields.

\section{Conclusion\label{conc}}

To conclude with, in this paper we have exposed an irreducible procedure for
approaching systems with second-class constraints reducible of order $L$.
Our strategy includes three main steps. First, we express the Dirac bracket
for the reducible system in terms of an invertible matrix. Second, we
establish the equality between this Dirac bracket and that corresponding to
the intermediate theory, based on the constraints (\ref{l9}). Third, we
prove that there exists an irreducible second-class constraint set
equivalent with (\ref{l9}) such that the corresponding Dirac brackets
coincide. These three steps enforce the fact that the fundamental Dirac
brackets with respect to the original variables derived within the
irreducible and original reducible settings coincide. Moreover, the newly
added variables do not affect the Dirac bracket, so the canonical approach
to the initial reducible system can be developed in terms of the Dirac
bracket corresponding to the irreducible theory. The general procedure was
exemplified on Abelian gauge-fixed $p$-form gauge fields. It is important to
mention that our procedure does not spoil other important symmetries of the
original system, such as spacetime locality of second-class field theories.

\section*{Acknowledgments}

This work has been supported in part by the contract 2-CEx-06-11-92/2006
with the Romanian Ministry of Education and Research (M.Ed.C.) and by the
European Commission FP6 program MRTN-CT-2004-005104. Two of the authors
(C.B. and S.O.S.) thank Marc Henneaux for his kind hospitality at Universit%
\'{e} Libre de Bruxelles, where this paper was completed, and acknowledge
useful discussions with Glenn Barnich.

\end{document}